\documentclass{iopart}

\expandafter\let\csname equation*\endcsname\relax
\expandafter\let\csname endequation*\endcsname\relax
\usepackage{amsmath,amssymb,amsthm}
\usepackage{graphicx}
\usepackage{dcolumn}
\usepackage{bbm}
\usepackage{subfigure}

\newcommand{\ba}{\begin{array}}
\newcommand{\ea}{\end{array}}
\newcommand{\bqa}{\begin{eqnarray}}
\newcommand{\eqa}{\end{eqnarray}}

\begin{document}

\title{A universal, memory-assisted entropic uncertainty relation}
\author{Z.-H. Ma,$^{1,2,+}$ C.-M. Yao,$^{3,4,+}$ Z.-H. Chen,$^{5}$ S. Severini$^{2}$, A. Serafini$^{4}$}
\address{$^1$Department of Mathematics, Shanghai Jiao-Tong University, Shanghai, 200240, P. R. China}
\address{$^2$Department of Computer Science, University College
London, Gower St., WC1E 6BT London, United Kingdom}
\address{$^3$Department of Physics and Electronic Science, Hunan University of Arts and Science, Changde, 415000, China}
\address{$^4$Department of Physics and Astronomy, University College
London, Gower St., WC1E 6BT London, United Kingdom}
\address{$^5$Department of Science, Zhijiang College, Zhejiang
University of Technology, Hangzhou, 310024,  China}
\address{$^{+}$ These two authors contributed equally to this work.}

\date{\today}

\begin{abstract}
We derive a new memory-assisted entropic uncertainty relation 
for non-degenerate Hermitian observables where both quantum correlations, in the form of conditional von Neumann entropy, 
and quantum discord between system and memory play an explicit role.
Our relation is `universal', in the sense that it does not depend on the specific observable, but only on properties of the quantum state.
We contrast such an uncertainty relation with previously known memory-assisted relations based on entanglement and correlations. 
Further, we present a detailed comparative study of entanglement- and discord-assisted entropic uncertainty relations for systems of two qubits -- one of which plays the role of the memory -- subject to several forms of independent quantum noise, in both Markovian and non-Markovian regimes. 
We thus show explicitly that, partly due to the ubiquity and inherent resilience of quantum discord, 
discord-tightened entropic uncertainty relations often offer a better estimate of the uncertainties in play.
\end{abstract}


\maketitle

\section{Entropic uncertainty relations}

One of the key aspects of quantum theory 
is that it is fundamentally impossible to know certain things, such as a particle's position and momentum, 
simultaneously with infinite precision. 
In fact, quantum mechanical uncertainty principles assert fundamental limits on the precision with which certain pairs of physical properties, such 
as position and momentum, may be simultaneously known.
Originally observed by Heisenberg \cite{Heisenberg}, the uncertainty principle is best known in the Robertson-Schr\"odinger form \cite{Robertson}
$$\Delta X\Delta Y\geq \frac{1}{2}|\langle[X,Y]\rangle|,$$
where $\Delta X$ ( $\Delta Y$) represents the standard deviation of the corresponding
observable $X (Y)$, formally represented by a Hermitian operator.

The entropic uncertainty relation for any two general observables was first given by Deutsch in terms of an information-theoretic model \cite{Deutsch}. Afterwards, an improved version was given by Kraus and then proved by Maassen and Uiffink \cite{Kraus}, which strengthen and generalizes Heisenberg's uncertainty relations, and can be written as follows:
$$H(X)+H(Y)\geq \log_2\left(\frac{1}{c}\right),$$
where $H$ is the Shannon entropy of the measured observable and $c$ quantifies the `complementarity' between the observables: 
$c=\max_{(i,j)} |\langle x_i | y_j\rangle|^2$ if $X$ and $Y$ are non-degenerate 
observables ($|x_i\rangle , |y_j\rangle$ being the eigenvectors of $X$ and $Y$, respectively).
Recently, the uncertainty principle in terms of entropy has been extended to the case involving entanglement with a quantum memory: It 
was proven by Berta et al.\ \cite{Berta} that
\begin{equation}
S(X|B)+S(Y|B)\geq \log_2\left(\frac{1}{c}\right)+S(A|B), \label{berta}
\end{equation}
where $S(X|B)=S\left[\sum_{j}\left(|x_j\rangle\langle x_j|\otimes\mathbbm{1}\right)\varrho_{AB}\left(|x_j\rangle\langle x_j|\otimes\mathbbm{1}\right)\right]$ and  
$(S(S|B))=S\left[\sum_{j}\left(|y_j\rangle\langle y_j|\otimes{1}\right)\varrho_{AB}\left(|y_j\rangle\langle y_j|\otimes{1}\right)\right]$ 
are the average conditional von Neumann entropies representing the uncertainty of the measurement outcomes of $X$ and $Y$ obtained using the information stored in system $B$, given that the initial system plus memory state was $\varrho_{AB}$. 
The quantity $S(A|B) = S(\varrho_{AB})-S({\rm Tr}_A(\varrho_{AB}))$ represents instead the conditional von Neumann entropy between system $A$ and $B$, defined in analogy with the classical definition of conditional entropy. The measurement outcomes of two incompatible observables on a particle can be precisely predicted when it is maximally entangled with a quantum memory, as indicated by Eq.(\ref{berta}). This entanglement-assisted entropic uncertainty relation was promptly experimentally tested [6].

Note that the quantum conditional entropy $S(A|B)$, appearing in the lower bound above, is a quantifier of quantum correlations that 
can be negative for entangled states, and thus may tighten 
the uncertainty relation without memory. Quantum correlations may also be assessed by quantum discord, which we define in the following. 
The total correlations between two quantum systems $A$ and $B$ are quantified by the quantum mutual information 
\begin{equation}\label{Total}
    \mathcal{I}(\varrho_{AB})=S(\varrho_{A})+S(\varrho_{B})-S(\varrho_{AB}) \; ,
\end{equation}
where $\varrho_{A(B)}= \mathrm{Tr}_{B(A)}(\varrho_{AB})$.
On the other hand, the classical part of correlations is defined as the maximum information that can be obtained by performing a local measurement,  which is defined as $\mathcal{I}(\varrho_{AB}|\{\hat{\Pi}_{A}^{j}\})=S(\varrho_{B})-\sum_j p_j S(\varrho_{B|j})$, where $\varrho_{B|j}=\tr_{A}((\hat{\Pi}_{A}^{j}\otimes \mathbbm{1})\varrho(\hat{\Pi}_{A}^{j}\otimes \mathbbm{1}))/\tr_{AB}[(\hat{\Pi}_{A}^{j}\otimes I)\varrho(\hat{\Pi}_{A}^{j}\otimes \mathbbm{1})]$,  $\{\hat{\Pi}_{A}^{j}\}$ are POVM locally performed  on subsystem $A$,  $p_j$ is the probability of the measurement outcome $j$.
Classical correlations are thus quantified by \cite{ved}: 
\begin{equation}\label{classical}J_{A}(\varrho_{AB})=\mathrm{sup}_{\{\hat{\Pi}_{A}^{j}\}}\mathcal{I}(\varrho_{AB}|\{\hat{\Pi}_{A}^{j}\})\end{equation}
Then  quantum $A$-side discord (quantum correlation) is defined as the difference of total correlation and $A-$side classical correlation \cite{Ollivier}:
\begin{equation}
    D_{A}(\varrho_{AB})=\mathcal{I}(\varrho_{AB})- J_{A}(\varrho_{AB}) \; .\label{Bdiscord}
\end{equation}


Recently, by considering the role of quantum discord and classical correlations of the joint system-memory state, Pati  et al.\ \cite{Pati} 
obtained a modified entropic uncertainty relation that tightens the lower bound (\ref{berta}) of Berta et al.\:
\begin{equation}
S(R|B)+S(S|B)\geq \log_{2}\frac{1}{c}+S(A|B)
+\max\{0, D_{A}(\varrho_{AB})- J_{A}(\varrho_{AB})\} \; .\label{pati}
\end{equation}

Entropic uncertainty relations have been shown to hold fundamental consequences for the security 
of cryptographic protocols \cite{tomamichel,ng,moloktov}, 
the foundations of thermodynamics \cite{hanggi} and entanglement theory \cite{guhne,niekamp}. 
See also \cite{wehner} for a quite recent review.

The main finding of the present work is the derivation of a new discord-assisted uncertainty relation, which we shall then contrast with the already known ones (Section 2). 
We will then move on to consider the behaviour of the different uncertainty relations under decoherence (Section 3), 
both to shed further light on our newly derived universal entropic uncertainty relation, and to provide the reader with a detailed comparative analysis of the different memory-assisted entropic uncertainty relations in realistic, noisy situations. We shall draw conclusions in Section 4.

\section{An observable-indipendent entropic uncertainty relation}

In the following, we use $X$ and $Z$ to denote two {\em non-degenerate} Hermitian observables described by the POVMs $X=\{X_{j}\}$ and $Z=\{Z_{k}\}$, 
where $X_j$ and $Z_{j}$ are orthogonal projectors. Otherwise, we shall apply all the notation introduced in the previous section.
Please also bear in mind that, in this paper, we always use $S(\varrho_{AB})$ ($S(X|B)$) to denote the von Neumann entropy (conditional von Neumann entropy) of the quantum state $\varrho_{AB}$ ($(X\otimes I)\varrho_{AB}$), while we use $H(X)$ to denote the Shannon entropy of the discrete probability distribution $P$ of measurement outcomes for $X$.
Further, we will make use of the following lemma: \medskip

\noindent {\bf Lemma 1.} Let $X:=\{X_{j}\}$ and $Z:=\{Z_{k}\}$ be arbitrary POVMs on $A$, then for all single-partite state $\varrho_{A}$,
\begin{equation}
 H(X)+ H(Z)\geq \log\frac{1}{c(X)}+\log\frac{1}{c(Z)}\\
 +2S(A),\label{Cond2}
\end{equation}
where  $c(X):=\max_{i} \tr(X_{i})$ and $c(Z):=\max_{i} \tr(Z_{i})$, $H(X)$ is the Shannon entropy of the probability $p_i :={\rm Tr}[(X_i ) \varrho_{A}]$, $H(Z)$ is the Shannon entropy of the probability $q_i :={\rm Tr}[(Z_i ) \varrho_{A}]$.\medskip

\noindent {\bf Proof.} From Corollary 7 of Ref.\ \cite{Coles}, we know that 
\begin{equation}
 H(X)\geq \log\frac{1}{c(X)}+S(A) \; .\label{Cond2a}
\end{equation}
By using the above relation twice, we get the result:
\begin{equation}
 H(X)+ H(Z)\geq \log\frac{1}{c(X)}+\log\frac{1}{c(Z)}\\
 +2S(A) \; , \label{Cond2}
\end{equation}
which proves our lemma. \qed\medskip

Clearly, if $X$ and $Z$ are non degenerate Hermitian observables, such that their POVM elements are all one-dimensional projectors, 
the inequality of Lemma 1 becomes:
\begin{equation}
 H(X)+ H(Z)\geq 
 2S(A) \; . \label{Cond3}
\end{equation}

We can now move on to our main result:\medskip

\noindent {\bf Theorem 1.} Let $X:=\{X_{j}\}$ and $Z:=\{Z_{k}\}$ be non-degenerate Hermitian observables on subsystem $A$, and 
let $\varrho_{AB}$ be any bipartite state of systems $A$ and $B$. One has
\begin{align}
 S(X|B)+ S(Z|B)\geq \nonumber\\
 +2S(A|B)+ 2D_{A}(\varrho_{AB})\label{bound} \; ,
\end{align}
where $D_{A}(\varrho_{AB})$ is the quantum discord, $S(X|B)$ and $S(Z|B)$ conditional entropies after measurements on $A$, and 
$S(A|B)$ is the von Neumann conditional entropy of state $\varrho_{AB}$, all defined above.\medskip

\noindent {\bf Proof.}
Consider a bipartite density operator $\varrho_{AB}$. If Alice performs a measurement of an observable $X$ on subsystem $A$, then the post-measurement state
is $\varrho_{AB}^X = \sum_i(X_{i} \otimes {\mathbbm 1}) \varrho_{AB} (X_i \otimes {\mathbbm 1}) = \sum_i p_i X_{i} \otimes \varrho_{B|i}$, where $p_i =
{\rm Tr}[(X_{i} \otimes I) \varrho_{AB}]$ is the probability of obtaining the $i^{\rm th}$ outcome and $\varrho_{B|i}={\rm Tr}_{A}[(X_{i} \otimes I) \varrho_{AB} (X_{i} \otimes {\mathbbm 1})]/p_i$ is
the conditional state of the memory $B$ corresponding to this outcome.
The conditional von Neumann entropy $S(X|B)$ denotes the ignorance about the
measurement outcome $X$ given information stored in a quantum memory held by an observer $B$. Thus, $S(X|B)=S(\varrho_{AB}^X)-S(\varrho_B)$ is the conditional entropy of the
state $\varrho_{AB}^X$. This is given by
\begin{equation}
S(X|B) = \sum_i p_i S(\varrho_B|i) + H(P) -S(\varrho_B)\label{Cond}
\end{equation}
Here $P:=(p_i)$ is the probability distribution of the outcomes.

It is worth noticing that,  if $B$ is a null system then, of the three terms in (\ref{Cond}), only the term $H(P)$ survives, {\em i.e.}\ $S(X|B) = H(P)$: the conditional von Neumann entropy of the quantum state after the measurement is the same as the Shannon entropy of the 
discrete probability distribution of measurement outcomes.

Now, denote by $P:=(p_i)$ the probability distribution of the measurement outcomes, so 
$p_i:={\rm Tr}[(X_i \otimes I) \varrho_{AB}]$; it is clear that ${\rm Tr}[(X_i \otimes I) \varrho_{AB}]={\rm Tr}[(X_i ) \varrho_{A}]$. Hence, the probability distribution $P:=(p_i)$ only depends on the reduced density matrix $\varrho_{A}$. Then the Shannon entropy of $P$ only depends on the single partite state $\varrho_{A}$.

For the bipartite state $\varrho_{AB}$, with Hermitian measurements $X:=\{X_{j}\}$ and $Z:=\{Z_{k}\}$ performed on  subsystem $A$, the following holds:
\begin{align}
&  S\left(X|B\right)  +S\left(Z|B\right) \nonumber\\
&  =H\left(X\right) -\mathcal{I}(\varrho_{AB}|\{X_{j}\})  +H\left(Z\right) -\mathcal{I}(\varrho_{AB}|\{Z_{k}\}) \nonumber\\
&  \geq H\left(X\right)  +H\left(Z|\right) -2J_{A}\left(\varrho_{AB}\right)\nonumber\\
&  \geq2S(\varrho_{A})-2J_{A}(\varrho_{AB}) \nonumber\\
&  =2S(A|B)+ 2D_{A}(\varrho_{AB})\label{proof-of-new-ineq} \, . %
\end{align}
The first identity is a consequence of Eq.~(\ref{Cond}). The
first inequality follows from the definition of the classical correlation
$J_{A}\left(  \varrho_{AB}\right)$. Since  $J_{A}\left(  \varrho_{AB}\right)$ is defined as the maximal among all POVMs for $\mathcal{I}(\varrho_{AB}|\{\hat{\Pi}_{A}^{j}\})$, 
in general, $\mathcal{I}(\varrho_{AB}|\{X_{j}\})\leq J_{A}\left(  \varrho_{AB}\right)$; similarly, $\mathcal{I}(\varrho_{AB}|\{Z_{k}\})\leq J_{A}\left(  \varrho_{AB}\right)$.
The second inequality comes from Eq.~(\ref{Cond3}), a consequence of the Lemma reported above. 
Final, the last equality follows by the definition of quantum discord. \qed\medskip

We now intend to compare our bound with the inequality (\ref{pati}), which is also related to quantum discord.
The term $\frac{1}{c}$ in (\ref{pati}) quantifies the compatibility of the two observables, and thus accounts for specific information 
concerning the measurements carried out. Thus, one should expect such a bound to be typically tighter than the relation (\ref{bound}). 
This is in fact often the case. However, aside from the intrinsic value of a universal, observable-independent relation, 
we find that, in several significant cases which we shall cover in the next section, our bound is more strict than the Inequality (\ref{pati}), and is almost the same as the actual value of the uncertainty. In a sense, in such cases, quantum correlations between the two subsystems make up for the absence of a measurement-specific term like $\frac{1}{c}$ of Ineq.~(\ref{pati}).

In particular, as shown in what follows, 
our bound turns out to improve quite often on 
Berta et al.'s lower bound of Inequality (\ref{berta}), based on quantum correlations.

\subsection{Examples}

Let us first illustrate the relevance of our relationship by considering some ad-hoc instances.

\subsubsection{Two-qubit Werner states.}
 
Consider the two-qubit Werner state
$\varrho_{AB}=\frac{1-f}{4}I_{A}\otimes I_{B}+f|\Psi^{-}\rangle\langle\Psi^{-}|$,
where $|\Psi^{-}\rangle=(|01\rangle-|10\rangle)/\sqrt{2}$ is the anti-symmetric Bell state, and $0\leq f\leq 1$.
We choose observables $X$ and $Z$ as the two spin observables $\sigma_{x}$ and $\sigma_{z}$. Then the uncertainty 
can be determined as:
\begin{align*}
S(X|B)+ S(Z|B)=2-(1-f)\log_2(1-f)-(1+f)\log_2(1+f) \; .
\end{align*}
The lower bound in (\ref{bound}) reads:
\begin{align*}
 2D_{A}(\varrho_{AB})
=2-(1-f)\log_2(1-f)-(1+f)\log_2(1+f) \; ,
\end{align*}
which is exactly equal to the uncertainty $S(X|B)+ S(Z|B)$.
this also coincides with the lower bound (\ref{pati}) of Ref.~\cite{Pati}:
\begin{align*}2-(1-f)\log_2(1-f)-(1+f)\log_2(1+f) \; .
\end{align*}
Instead, the lower bound (\ref{berta}) of Ref.~\cite{Berta} is
\begin{align*}2-\frac{1+3f}{4}\log_2((1+3f))-\frac{3(1-f)}{4}\log_2(1-f)\; ,\end{align*} 
which is smaller, and thus less informative, than the other two.

\subsubsection{Two-qutrit Werner states.} 
For two qutits, a Werner state can be written as 
$\varrho_{AB}=\frac{1-f}{6}\Pi^{+}+\frac{f}{3}\Pi^{-}$, 
where $\Pi^{+}$ is the projector onto the symmetric subspace
and $\Pi^{-}$ is the projector onto the antisymmetric subspace.
We choose observables $X$ and $Z$ as two generators of $SU(3)$ and define :
\begin{align*}
|0\rangle=(1,0,0)^{T},
|1\rangle=(0,1,0)^{T},|2\rangle=(0,0,1)^{T} \, ,
\end{align*}
\begin{align*}
X=|0\rangle\langle 1|+|1\rangle\langle 0|, \quad 
Z=|0\rangle\langle 0|-|1\rangle\langle 1| \, ,
\end{align*}
such that
\begin{align*}
S(X|B)+ S(Z|B)
=f+3-(1-f)\log_2(1-f)-(1+f)\log_2(1+f) \; ,
\end{align*}
\begin{align*}
D_{A}(\varrho_{AB})&=2+f\log_2(\frac{f}{2})+(1-f)\log_2(\frac{1-f}{4})\\&
-\frac{1-f}{2}\log_2(1-f)-\frac{1+f}{2}\log_2(\frac{1+f}{2}).
\end{align*}
Our bound (\ref{bound}) then reads:
\begin{align*}
2S(A|B)+ 2D_{A}(\varrho_{AB})\\=f+3-(1-f)\log_2(1-f)-(1+f)\log_2(1+f)\; ,
\end{align*}
which is equal to the uncertainty $S(X|B)+ S(Z|B)$
But the lower bound (\ref{pati}) of Ref.~\cite{Pati} is
\begin{align*}
-f+1-(1-f)\log_2(1-f)-f\log_2f\\+
\max\{0,2f+f\log_2f-\log_2\frac{3(1+f)}{4}\\-f\log_2(1+f)\}\\
=f+3-\log_23-(1-f)\log_2(1-f)-(1+f)\log_2(1+f)\; ,
\end{align*} which is smaller than what we obtained.
For the lower bound (\ref{berta}) of Ref.~\cite{Berta} one has
\begin{align*}
-f+1-(1-f)\log_2(1-f)-f\log_2f \; ,
\end{align*} which is smaller than our lower bound. 
We have thus identified a situation, with 2-qutrit Werner states, {\em where our bound performs better than the previously known ones}.

\subsubsection{Isotropic states.}  
Consider a bipartite isotropic state with local Hilbert space dimension $d$, $\varrho=f \phi_{d}+\frac{1-f}{d^2-1}(I-\phi_{d})$.
For $d=2$, consider the following observables
\begin{align*}
|0\rangle=(1,0)^{T},|1\rangle=(0,1)^{T},
\end{align*}
\begin{align*}
X=|0\rangle\langle 1|+|1\rangle\langle 0|,Z=|0\rangle\langle 0|-|1\rangle\langle 1|.
\end{align*}
Then
\begin{align*}
S(X|B)+ S(Z|B)=-\frac{2}{3}(-2f+2(1-f)\log_2(1-f)+\log_2\frac{4(1+2f)}{27}+2f\log_2(1+2f)) \, ,
\end{align*}
\begin{align*}
D_{A}(\varrho_{AB})=\frac{1-f}{3}\log_2(\frac{1-f}{3})+f\log_2f-\frac{1+2f}{3}\log_2\frac{1+2f}{6}.
\end{align*}
The universal lower bound (\ref{bound}) reads:
\begin{align*}
-\frac{2}{3}(-2f+2(1-f)\log_2(1-f)+\log_2\frac{4(1+2f)}{27}+2f\log_2(1+2f)) \, ,
\end{align*}
which is exactly equal to the uncertainty $S(X|B)+ S(Z|B)$.
The lower bound (\ref{pati}) of Ref.~\cite{Pati} is also equal to the uncertainty $S(X|B)+ S(Z|B)$ in this case.
However, the lower bound (\ref{berta}) of Ref.~\cite{Berta} equals 
\begin{align*}
-(1-f)\log_2\frac{1-f}{3}-f\log_2 f \; ,
\end{align*}
which is always smaller than what obtained with discord-assisted bounds. 

\begin{figure}[t!]
\centering
\includegraphics[width=0.4\textwidth]{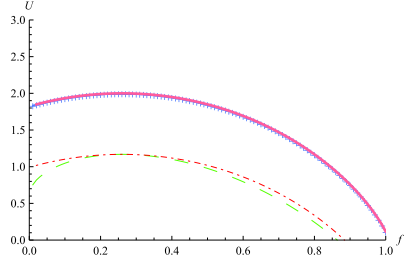}
\caption{Plot of the different lower bounds for isotropic states with dimension $d=2$ 
and 
observables given by the matrices in Eqs.~(\ref{X1},\ref{Z1}). 
The upper blue `+'line and pink solid lines represent the uncertainty $S(X|B)+ S(Z|B)$ and the universal lower bound (\ref{bound}),
the lower green `-' line and red `.-' line represent the results of the bounds (\ref{berta}) pof Ref.~\cite{Berta} and (\ref{pati}) 
of Ref.~\cite{Pati}.\label{bound1}}
\end{figure}

When $d=3$, instead, by choosing the observables
\begin{align*}
|0\rangle=(1,0,0)^{T},|1\rangle=(0,1,0)^{T},|2\rangle=(0,0,1)^{T},
\end{align*}
\begin{align*}
X=|0\rangle\langle 1|+|1\rangle\langle 0|,Z=|0\rangle\langle 0|-|1\rangle\langle 1| ,
\end{align*}
one gets
\begin{align*}
S(X|B)+ S(Z|B)=-\frac{1}{2}(3(1-f)\log_2\frac{1-f}{8}+\log_2\frac{27(1+3f)}{4}+3f\log_2\frac{1+3f}{12}) ,
\end{align*}
\begin{align*}
D_{A}(\varrho_{AB})=\frac{1-f}{4}\log_2(\frac{1-f}{8})+f\log_2f-\frac{1+3f}{4}\log_2\frac{1+3f}{12}.
\end{align*}
The universal lower bound (\ref{bound}) then reads:
\begin{align*}
-\frac{1}{2}(3(1-f)\log_2\frac{1-f}{8}+\log_2\frac{27(1+3f)}{4}+3f\log_2\frac{1+3f}{12}),
\end{align*}
which is equal to the uncertainty $S(X|B)+ S(Z|B)$.
The lower bound (\ref{pati}) of Ref.~\cite{Pati} is instead
\begin{align*}
-\frac{1}{2}(3(1-f)\log_2\frac{1-f}{8}+\log_2\frac{243(1+3f)}{4}+3f\log_2\frac{1+3f}{12}),
\end{align*}
which is smaller than ours.
The lower bound (\ref{berta}) of Ref.~\cite{Berta} is equal to
\begin{align*}
-(1-f)\log_2(1-f)-f\log_2 f-(3f-3)-\log_23 \; ,
\end{align*}
which is also smaller than our bound.
\begin{figure}[t!]
\centering
\includegraphics[width=0.4\textwidth]{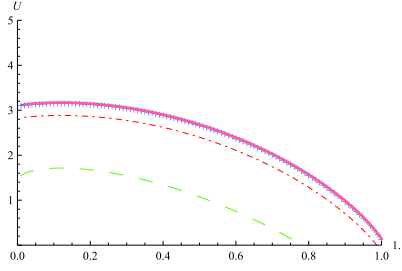}
\caption{Plot of the different lower bounds for isotropic states with dimension $d=3$ and 
observables given by the matrices in Eqs.~(\ref{X2},\ref{Z2}). 
The upper blue `+'line and pink solid lines represent the uncertainty $S(X|B)+ S(Z|B)$ and the universal lower bound (\ref{bound}),
the lower green `-' line and red `.-' line represent the results of the bounds (\ref{berta}) pof Ref.~\cite{Berta} and (\ref{pati}) 
of Ref.~\cite{Pati}. \label{bound2}}
\end{figure}

For an isotropic state with $d=2$, other cases can be constructed where our lower bound provides one with a substantial 
advantage, such as
\begin{equation}
X=\begin{pmatrix}
0.272007 , 0.0483473+0.584816 i \\
0.0483473-0.584816 i, 0.246297
\end{pmatrix},\label{X1}
\end{equation}
\begin{equation}
Z=\begin{pmatrix}
0.43916 , 0.857154+0.976248 i \\
0.857154-0.976248 i, 0.515329
\end{pmatrix}. \label{Z1}
\end{equation}
The different bounds for the observables above are displayed in Fig.~\ref{bound1}.
Likewise, cases where our lower bound is tighter can be found for $d=3$, such as
\begin{equation}
X=\begin{pmatrix}
0.246301 , 0.267394 + 0.627628 i,0.155311 + 0.270053 i \\
0.267394 - 0.627628 i, 0.752065,0.231887 + 0.500147 i\\
0.155311 - 0.270053 i,0.231887 - 0.500147 i,0.94377
\end{pmatrix}, \label{X2}
\end{equation}
\begin{equation}
Z=\begin{pmatrix}
0.586665 , 0.146795 + 0.957852 i,0.687252 + 0.677623 i \\
0.146795 - 0.957852 i, 0.709581,0.405322 + 0.525615 i\\
0.687252 - 0.677623 i,0.405322 - 0.525615 i,0.901804
\end{pmatrix}, \label{Z2}
\end{equation}
whose uncertainties and different bounds are depicted in Fig.~\ref{bound2}

\subsubsection{Qubit-qudit states.} 
\begin{figure}[t!]
\centering
\includegraphics[width=0.4\textwidth]{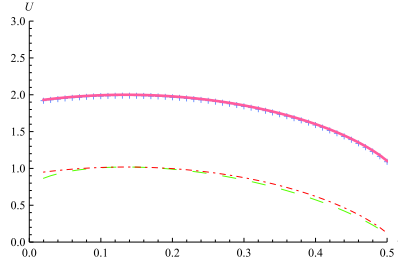}
\caption{Plot of the different lower bounds for the qubit-qutrit state defined in the text and 
observables given by the matrices in Eqs.~(\ref{X3},\ref{Z3}). 
The upper blue `+'line and pink solid lines represent the uncertainty $S(X|B)+ S(Z|B)$ and the universal lower bound (\ref{bound}),
the lower green `-' line and red `.-' line represent the results of the bounds (\ref{berta}) pof Ref.~\cite{Berta} and (\ref{pati}) 
of Ref.~\cite{Pati}. \label{bound3}}
\end{figure}

Consider the qubit-qutrit state 
$\varrho=\alpha (|02\rangle\langle 02|+|12\rangle\langle 12|)+\beta (|\phi^{+}\rangle\langle\phi^{+}|+|\phi^{-}\rangle\langle\phi^{-}|+
|\psi^{+}\rangle\langle\psi^{+}|)+\gamma|\psi^{-}\rangle\langle\psi^{-}|$, 
where 
$\phi_{\pm}=\frac{1}{\sqrt{2}}(|00\rangle\pm|11\rangle)$ and $\psi_{\pm}=\frac{1}{\sqrt{2}}(|01\rangle\pm|10\rangle))$,
as well as the following observables:
\begin{align*}
|0\rangle=(1,0)^{T},|1\rangle=(0,1)^{T},
\end{align*}
\begin{align*}
X=|0\rangle\langle 1|+|1\rangle\langle 0|,Z=|0\rangle\langle 0|-|1\rangle\langle 1|.
\end{align*}
Then 
\begin{align*}
S(X|B)+ S(Z|B)=4\alpha-4\beta-4\beta \log_2(\beta)-2(\beta+\gamma)\log_2(\beta+\gamma)+2(3\beta+\gamma)\log_2(3\beta+\gamma) \, .
\end{align*}
Our bound (\ref{bound}) and the lower bound (\ref{pati}) of Ref.~\cite{Pati} are all equal to the uncertainty $S(X|B)+ S(Z|B)$
in this instance.
The bound (\ref{berta}) of Ref.~\cite{Berta} is instead
\begin{align*}
4\alpha-3\beta \log_2(\beta)-\gamma \log_2(\gamma)+(3\beta+\gamma)\log_2(3\beta+\gamma) \; ,
\end{align*}
which is always smaller than the previous bounds.
Under the following choice of observables:
\begin{equation}
X=\begin{pmatrix}
0.826411 , 0.443371+0.745704 i \\
0.443371-0.745704 i, 0.459166
\end{pmatrix}, \label{X3}
\end{equation}
\begin{equation}
Z=\begin{pmatrix}
0.832848 , 0.191194+0.608568 i \\
0.191194-0.608568 i, 0.509301
\end{pmatrix}, \label{Z3}
\end{equation}
and parameters $\alpha=0.25$, $0\le\gamma\le0.5$ then, as depicted in Fig.~\ref{bound3}, 
the universal bound we derived is tighter than any of the previously known ones.

\begin{figure}[t!]
\centering
\includegraphics[width=0.4\textwidth]{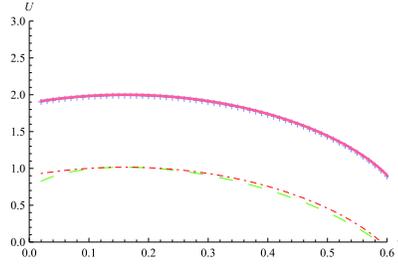}
\caption{Plot of the different lower bounds for the qubit by four-dimensional system state defined in the main text and 
observables given by the matrices in Eqs.~(\ref{X4},\ref{Z4}). 
The upper blue `+'line and pink solid lines represent the uncertainty $S(X|B)+ S(Z|B)$ and the universal lower bound (\ref{bound}),
the lower green `-' line and red `.-' line represent the results of the bounds (\ref{berta}) pof Ref.~\cite{Berta} and (\ref{pati}) 
of Ref.~\cite{Pati}. \label{bound4}}
\end{figure}

For the following state of one qubit times a four-dimensional quantum system
$\varrho=\alpha (|02\rangle\langle 02|+|03\rangle\langle 03|+|12\rangle\langle 12|+|13\rangle\langle 13|)+\beta (|\phi^{+}\rangle\langle\phi^{+}|+|\phi^{-}\rangle\langle\phi^{-}|+
|\psi^{+}\rangle\langle\psi^{+}|)+\gamma|\psi^{-}\rangle\langle\psi^{-}|$, with observables
\begin{align*}
|0\rangle=(1,0,0)^{T},|1\rangle=(0,1,0)^{T},|2\rangle=(0,0,1)^{T},
\end{align*}
\begin{align*}
X=|0\rangle\langle 1|+|1\rangle\langle 0|,
Z=|0\rangle\langle 0|-|1\rangle\langle 1|,
\end{align*}
one has
\begin{align*}
S(X|B)+ S(Z|B)=8\alpha-4\beta-4\beta \log_2(\beta)-2(\beta+\gamma)\log_2(\beta+\gamma)+2(3\beta+\gamma)\log_2(3\beta+\gamma).
\end{align*}
The universal bound (\ref{bound}) and the lower bound (\ref{pati}) of Ref.~\cite{Pati} are all equal to the uncertainty $S(X|B)+ S(Z|B)$,
while the bound (\ref{berta}) of Ref.~\cite{Berta} is always smaller and reads
\begin{align*}
8\alpha-3\beta \log_2(\beta)-\gamma \log_2(\gamma)+(3\beta+\gamma)\log_2(3\beta+\gamma).
\end{align*}
Under the following choice of observables:
\begin{equation}
X=\begin{pmatrix}
0.370786 , 0.344509+0.694499 i \\
0.344509-0.694499 i, 0.60978
\end{pmatrix},\label{X4}
\end{equation}
\begin{equation}
Z=\begin{pmatrix}
0.303997 , 0.332044+0.448198 i \\
0.332044-0.448198 i, 0.342387
\end{pmatrix}, \label{Z4}
\end{equation}
and parameters $\alpha=0.1$, $0\le\gamma\le0.6$ then, as shown in Fig.~\ref{bound4}, 
the universal bound we derived is tighter than any of the previously known ones.

\section{Uncertainty relations under decoherence}

In the real world, as already remarked, quantum states are unavoidably disturbed by decoherence induced by the environment. The extent to which the environment affects quantum entanglement or quantum and classical correlations beyond entanglement is a central problem, and much work has been done along these lines \cite{T. Yu,maziero}, in both Markovian \cite{Francesco} and non-Markovian \cite{Wang} regimes. It is thus natural to ask what impact such environmental decoherence has on the quantities entering entropic uncertainty relations in the presence of a quantum memory. Recently, Z. Y. Xu et al.\ \cite{Z. Y.} considered the behaviour of the uncertainty relation under the action of local unital and non-unital noisy channels. Thus, they found out that while unital noise increases the amount of uncertainty, the amplitude- damping nonunital noises may reduce the amount of uncertainty, and bring it closer to its lower bound in the long-time limit. These results shed light on the different competitive mechanisms  governing quantum correlations on the one hand and the minimal missing information after local measurements on the other.
 
In this section, we focus on two quantum bits, with one of the two qubits acting as a memory, and examine the behaviour of 
different entropic uncertainty relations -- including the newly derived one of Theorem 1 -- with assisting quantum and classical correlations when the two qubits interact with independent environments, in both Markovian and non-Markovian regimes. The most common noise channels (amplitude and phase damping) are analysed. Here, we shall focus on three scenarios in succession: first, we discuss the influence of system-reservoir dynamics of quantum and classical correlations on the entropic uncertainty relation, and compare the differences among three entropic uncertainty relations in Eq. (\ref{berta}), (\ref{pati}) and (\ref{bound}); second, we explore non-Markovian dynamics influence on the entropic uncertainty relation; and third, we discuss two special examples.

\subsection{Uncertainty relations under unital and non-unital local noisy channels}

In order to investigate the behaviour of the uncertainty relation under the influence of independent local noisy channels, 
in what follows we will consider a system $S$ comprised of qubits $A$ and $B$, each of them interacting independently 
with its own environment $E_A$ and $E_B$, respectively. The dynamics of two qubits interacting independently with individual environments are described by the solutions of the appropriate Born-Markov Lindblad equations \cite{H. P.}, which can be described conveniently in the Kraus operator formalism \cite{M. A. Nielsen}. Given an initial state for two qubits, its evolution can be written compactly as
\begin{equation}
\varrho_{AB}(t)=\sum\limits_{uv}M_{uv}\varrho_{AB}(0)M_{uv}^{\dag} ,
\end{equation}
where the Kraus operators $M_{u,v}=M_u\otimes M_v$ \cite{M. A. Nielsen} satisfy the completeness relation 
$\sum\limits_{u,v}M_{u,v}^{\dag}  M_{u,v}=\mathbbm{1}$ at all times. 
The operators $M_u$ and $M_v$ describe the one-qubit quantum channels. 

In the following we shall restrict to two-qubit states with maximally mixed local states that can be written in the form
\begin{equation}
\varrho_{AB} = \frac14\left(\mathbbm{1}+\sum\limits_{i=1}^3(c_{i}\sigma_i^A\otimes\sigma_i^{B}) \right) ,
\end{equation}
where $\sigma_i^R$ is the standard Pauli matrix in direction $j$ acting on the space of subsystem $R$ for $R=A,B$ and  
$c_{i}$ is a triple of real coefficients satisfying $0\le c_i \le 1$. 
Including environments, the whole initial state will be taken as $\varrho_{AB}\otimes|00\rangle_{E_A E_B}$, 
where $|00\rangle_{E_A E_B}$ is the nominal vacuum state of the environments 
$E_A$ and $E_B$ in which the qubits $A$ and $B$, respectively, are immersed. 
We present below what happens to the entropic uncertainty relation 
for some qubit channels of broad interest ({\em i.e.}, amplitude damping and phase damping).

\subsubsection{Amplitude damping channel}

The amplitude-damping channel, which is a classical noise process representing the dissipative interaction between 
system $S$ and the environment $E$, can be modelled by treating $E$ as a large collection of independent harmonic oscillators interacting weakly with $S$ \cite{H. P.}. The effect of a dissipative channel over one qubit is depicted by the following map
\begin{equation}
\begin{split}
\vert 0 \rangle_{S}\vert 0 \rangle_{E}\mapsto \vert 0 \rangle_{S}\vert 0 \rangle_{E},\\
\vert 1 \rangle_{S}\vert 0 \rangle_{E}\mapsto \sqrt{1-p}\vert 1 \rangle_{S}\vert 0 \rangle_{E}+\sqrt{p}\vert 0 \rangle_{S}\vert 1 \rangle_{E},
\end{split}
\end{equation}
where $|0\rangle_S$ is the ground and $|1\rangle_S$ the excited state of the qubit. 
The states $|0\rangle_E$ and $|1\rangle_E$ describe the states of the environment with no excitation and one excitation 
distributed over all its modes ({\em i.e.}, in the normal mode coupled to the qubit). 
The quantity $p\in[0,1]$ represents a decay probability, which will be a decreasing exponential function 
of time under Markov approximation. 
The corresponding Kraus operators describing the amplitude-damping channel acting on the system are given by \cite{M. A. Nielsen}
\begin{equation}
\begin{split}
M_{0}=\left(
\begin{array}{cc}
1 & 0  \\
0 & \sqrt{1-p} \\%
\end{array}%
\right)\otimes
\left(
\begin{array}{cc}
1 & 0  \\
0 & \sqrt{1-p} \\%
\end{array}%
\right),\\
M_{1}=\left(
\begin{array}{cc}
0 & \sqrt{p}  \\
0 & 0 \\%
\end{array}%
\right)\otimes
\left(
\begin{array}{cc}
0 & \sqrt{p}  \\
0 & 0 \\%
\end{array}
\right),
\end{split}
\end{equation}
The total system evolves under the action of the operators in Eq.~(7), 
obtained by tracing out the degrees of freedom of the reservoir, 
in the computational basis ${|00\rangle_{AB} , |01\rangle_{AB} , |10\rangle_{AB} , |11\rangle_{AB}}$ for qubits $A$ and $B$. 
The density operator after the action of the channel is given by \cite{J. Maziero}
\begin{equation}
\hspace*{-2cm}
\varrho_{AB}=\frac{1}{4}\left(
\begin{array}{cccc}
(1+p)^2+(1-p)^2c_3 & 0&0&(1-p)(c_1-c_2)  \\
0 &((1-c_3)+(1+c_3)p)(1-p)&(1-p)(c_1+c_2)&0  \\
0 &(1-p)(c_1+c_2)&((1-c_3)+(1+c_3)p)(1-p)&0  \\
(1-p)(c_1-c_2) &0 &0&(1-p)^2(1+c_3)  \\%
\end{array}%
\right) .
\end{equation}
Due to the X structure of the density matrices in Eq.(8), there is a simple closed expression for the concurrence $Con$ present in all bipartitions \cite{S. Luo}
\begin{equation}
Con(p)=2\max\{0,\lambda_1(p),\lambda_2(p)\},
\end{equation}
with $\lambda_1 (p)=|\varrho_{14} |-\sqrt{\varrho_{22} \varrho_{33} }$ and $\lambda_2 (p)=|\varrho_23 |-\sqrt(\varrho_11 \varrho_44 )$. 
We can also derive analytical expressions for mutual information and classical correlation:
\begin{eqnarray}
\fl I[\varrho_{AB}(p)]=-(1-p)\log_2(1-p)-(1+p)\log_2(1+p)\nonumber\\
+\frac{1}{4}(1-p)(1+c_1+c_2-c_3+p+c_3p) \log_2[(1-p)(1+c_1+c_2-c_3+p+c_3p)]\nonumber\\
+\frac{1}{4}(1-p)(1-c_1-c_2-c_3+p+c_3p) \log_2[(1-p)(1-c_1-c_2-c_3+p+c_3p)]\nonumber\\
+\frac{1}{4}(1+p^2+c_3(1-p)^2-\sqrt{(c_1-c_2)^2(1-p)^2+4p^2})\nonumber\\ \log_2[1+p^2+c_3(1-p)^2-\sqrt{(c_1-c_2)^2(1-p)^2+4p^2}]\nonumber\\
+\frac{1}{4}(1+p^2+c_3(1-p)^2+\sqrt{(c_1-c_2)^2(1-p)^2+4p^2})\nonumber\\ \log_2[1+p^2+c_3(1-p)^2+\sqrt{(c_1-c_2)^2(1-p)^2+4p^2}],\nonumber
\end{eqnarray}

\begin{eqnarray}
\fl C[\varrho_{AB}(p)]=\frac{1}{4}(1+p^2+c_3(1-p)^2-\sqrt{(c_1-c_2)^2(1-p)^2+4p^2}) \log_2[1+p^2+c_3(1-p)^2 \nonumber\\
-\sqrt{(c_1-c_2)^2(1-p)^2+4p^2}]+\frac{1}{4}(1+p^2+c_3(1-p)^2\nonumber\\
+\sqrt{(c_1-c_2)^2(1-p)^2+4p^2}) \log_2[1+p^2+c_3(1-p)^2 +\sqrt{(c_1-c_2)^2(1-p)^2+4p^2}] \nonumber\\-
\frac{1}{4}(1+p^2+c_3(1-p)^2+(c_1-c_2)(1-p)) \log_2[1+p^2+c_3(1-p)^2+(c_1-c_2)(1-p)]\nonumber\\-
\frac{1}{4}(1+p^2+c_3(1-p)^2-(c_1-c_2)(1-p)) \log_2[1+p^2+c_3(1-p)^2-(c_1-c_2)(1-p)] \nonumber\\
+\frac{1+c}{2}\log_2(1+c)+\frac{1-c}{2}\log_2(1-c),\nonumber
\end{eqnarray}
where $c=\max\{|c_1(1-p)|,|c_2(1-p)|,c_3(1-p)^2+p^2\}$.

The quantum discord is then given by [7]
\begin{equation}
D[\varrho_{AB}(p)]=I[\varrho_{AB}(p)]-C[\varrho_{AB}(p)] .
\end{equation}

If one chooses two of the Pauli observables $R=\sigma_i$ and $S=\sigma_j$ ($i, j=1, 2, 3$) as measurements, 
the left-hand side of Eq.~(\ref{berta}) can be written as
\begin{eqnarray}
\fl U=2+(1-p)\log_2(1-p)+(1+p)\log_2(1+p) \nonumber\\
-\frac{1}{2}[(1-\sqrt{c_1^2(1-p)^2+p^2})\log_2(1-\sqrt{c_1^2(1-p)^2+p^2})\nonumber\\+
(1+\sqrt{c_1^2(1-p)^2+p^2}) \log_2(1+\sqrt{c_1^2(1-p)^2+p^2})\nonumber\\+
(1-\sqrt{c_2^2(1-p)^2+p^2}) \log_2(1-\sqrt{c_2^2(1-p)^2+p^2})\nonumber\\+
(1+\sqrt{c_2^2(1-p)^2+p^2}) \log_2(1+\sqrt{c_2^2(1-p)^2+p^2}] . \nonumber
\end{eqnarray}

On the other hand, the complementarity c of the observables $\sigma_i$ and $\sigma_j$ is always equal to $1/2$, 
so that the right-hand sides of Eq. (\ref{berta}), Eq. (\ref{pati}) and Eq. (\ref{bound}), which we shall denote by $U_{b1}$, $U_{b2}$ and $U_{b3}$ 
take the form, respectively,
\begin{equation}
U_{b1}=1+S(\varrho_{AB})-S(\varrho_{B}) ,
\end{equation}
where $S(\varrho_{AB})=-\frac{1}{4}(1-p)(1+c_1+c_2-c_3+p+c_3p)\log_2[(1-p)(1+c_1+c_2-c_3+p+c_3p)]
-\frac{1}{4}(1-p)(1-c_1-c_2-c_3+p+c_3p)\log_2[(1-p)(1-c_1-c_2-c_3+p+c_3p)]
-\frac{1}{4}(1+p^2+c_3(1-p)^2-\sqrt{(c_1-c_2)^2(1-p)^2+4p^2})\log_2[(1+p^2+c_3(1-p)^2-\sqrt{(c_1-c_2)^2(1-p)^2+4p^2})]
-\frac{1}{4}(1+p^2+c_3(1-p)^2+\sqrt{(c_1-c_2)^2(1-p)^2+4p^2})\log_2[(1+p^2+c_3(1-p)^2+\sqrt{(c_1-c_2)^2(1-p)^2+4p^2})]$ ,
$S(\varrho_{B})=1-\frac{1-p}{2}\log_2(1-p)-\frac{1+p}{2}\log_2(1+p)$, 
\begin{equation}
U_{b2}=1+S(\varrho_{AB})-S(\varrho_{B})+\max\{0,D[\varrho_{AB}(p)]-C[\varrho_{AB}(p)]\} ,
\end{equation}
\begin{equation}
U_{b3}=2S(\varrho_{AB})-2S(\varrho_{B})+2 D[\varrho_{AB}(p)] .
\end{equation}

\begin{figure}
\subfigure[]
{ \begin{minipage}[b]{0.5\textwidth}
\includegraphics[width=1\textwidth]{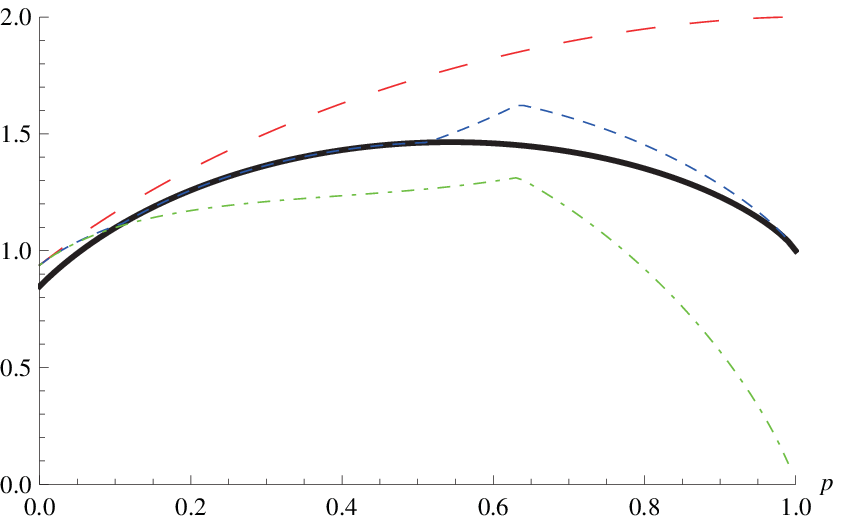}
\end{minipage} }
\subfigure[]
{ \begin{minipage}[b]{0.5\textwidth}
\includegraphics[width=1\textwidth]{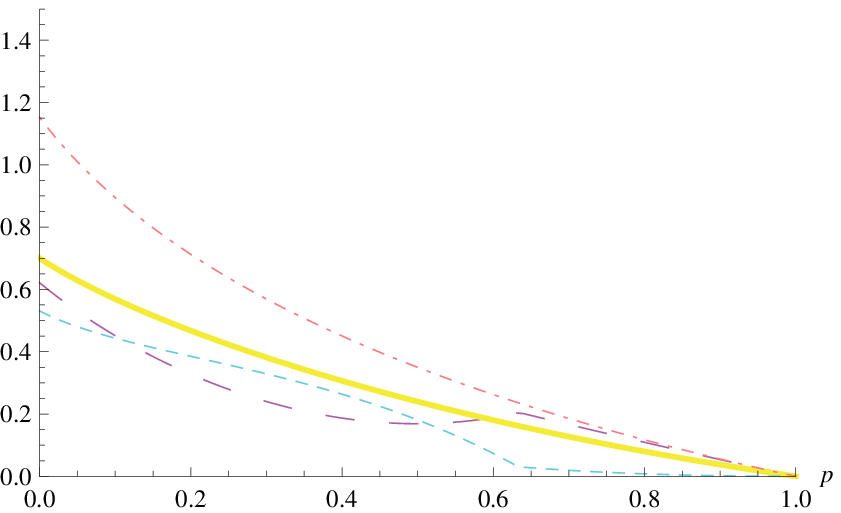}
\end{minipage} }
 \caption{(a) $U$ and $U_{bi} (i=1, 2, 3)$ of the observables $\sigma_1, \sigma_2$ with initial state $c_1=c_2=c_3=-0.8$ under local amplitude-damping channel with damping probability $p$. 
 $U$-red large dashing color, $U_{b1}$-black solid color, $U_{b2}$-blue dot color, 
 $U_{b3}$-green dot dashed color. 
(b) Concurrence ($Con$)-yellow solid color, discord (D)- magenta large dashing color, 
 classical correlation ($C$)-light blue dot color,  and mutual information (M)-pink dot dashed color.}
\label{fig:Fig1}
\end{figure}

Let us now choose the initial state $c_1=c_2=c_3=-0.8$. The specifics of the dynamics of the uncertainties clearly depend on the initial state, 
but these values represent a typical situation. 
Also, let us denote by $U$ the left hand side of Eqs.~(\ref{berta}), (\ref{pati}) and (\ref{bound}) when two Pauli observables $\sigma_1$ and $\sigma_2$
are chosen. 
As shown in fig. 1, the quantity $U$ will increase over time due to the gradual decay of correlations between system qubit $A$ 
and memory qubit $B$, while $U_{b1}$, $U_{b2}$ and $U_{b3}$, corresponding to the lower bounds of the uncertainty in Eq. (\ref{berta}), (\ref{pati}) and (\ref{bound}) respectively, 
will increase at first, and then decrease to asymptotic values. The dynamics of discord, entanglement and classical correlations present in qubits A and B are shown in the inset of Fig.\ 1. In this case, the non-unital channel induces disentanglement in asymptotic time; the quantum discord first decreases then increases for a short interval, and finally decreases to disappear asymptotically (if one assumes $p$ to 
fall exponentially in time). This dynamics induces substantial differences among the behaviours of $U_{b1}$, $U_{b2}$ and $U_{b3}$:
at very short times, $U_{b3}$ better approximates the evolution of the Shannon entropies, thanks to the permanence of classical correlations 
at such times, while $U_{b3}$ becomes tighter at intermediate times.

\subsubsection{Phase damping channel}

The phase-damping channel is a unital channel (where a maximally mixed input state is left unchanged) 
leads to a loss of quantum coherence without loss of energy. The map of this channel on a one-qubit system is given by
\begin{equation}
\begin{split}
\vert 0 \rangle_{S}\vert 0 \rangle_{E}\mapsto \vert 0 \rangle_{S}\vert 0 \rangle_{E},\\
\vert 1 \rangle_{S}\vert 0 \rangle_{E}\mapsto \sqrt{1-p}\vert 1 \rangle_{S}\vert 0 \rangle_{E}+\sqrt{p}\vert 1 \rangle_{S}\vert 1 \rangle_{E},
\end{split} .
\end{equation}
The corresponding Kraus operators describing the phase-damping channel for the system of qubits A and B can be written as
\begin{equation}
\begin{split}
M_{AB}^0=\left(
\begin{array}{cc}
1 & 0  \\
0 & \sqrt{1-p} \\%
\end{array}%
\right)\otimes
\left(
\begin{array}{cc}
1 & 0  \\
0 & \sqrt{1-p} \\%
\end{array}%
\right),
\\
M_{AB}^1=\left(
\begin{array}{cc}
0 & 0 \\
0 & \sqrt{p} \\%
\end{array}%
\right)\otimes
\left(
\begin{array}{cc}
0 & 0  \\
0 & \sqrt{p} \\%
\end{array}
\right),
\end{split} .
\end{equation}
For the initial state (5), the evolved density operator of the system $AB$, 
obtained by tracing out the degrees of freedom of the reservoirs, is given by
\begin{equation}
\varrho_{AB}=\frac{1}{4}\left(
\begin{array}{cccc}
\frac{1+c_3}{4} & 0&0&\frac{(1-p)c^{-}}{4}  \\
0 &\frac{1-c_3}{4}&\frac{(1-p)c^{+}}{4}&0  \\
0 &\frac{(1-p)c^{+}}{4}&\frac{1-c_3}{4}&0  \\
\frac{(1-p)c^{-}}{4} &0 &0&\frac{1+c_3}{4}  \\%
\end{array}%
\right) ,
\end{equation}
where $c^{\pm}=c_1\pm c_2$. 
The mutual information and the classical correlation present in qubits $A$ and $B$ can be computed analytically and are given by
\begin{equation}
\begin{split}
I[\varrho_{AB}(p)]=\frac{1}{4}(1+c_1+c_2-c_3-c_1p-c_2p) \log_2(1+c_1+c_2-c_3-c_1p-c_2p)\\
+\frac{1}{4}(1-c_1+c_2+c_3+c_1p-c_2p) \log_2(1-c_1+c_2+c_3+c_1p-c_2p)\\
+\frac{1}{4}(1+c_1-c_2+c_3-c_1p+c_2p) \log_2(1+c_1-c_2+c_3-c_1p+c_2p)\\
+\frac{1}{4}(1-c_1-c_2-c_3+c_1p+c_2p) \log_2(1-c_1-c_2-c_3+c_1p-c_2p),\\
\end{split} 
\end{equation}
\begin{equation}
C[\varrho_{AB}(p)]=\frac{1-c}{2}\log_2(1-c)+\frac{1+c}{2}\log_2(1+c) ,
\end{equation}
where $c=\max\{|c_1(1-p)|,|c_2(1-p)|,|c_3|\}$
The concurrence of qubits $A$ and $B$ is instead given by
\begin{equation}
\begin{split}
Con(p)=\frac{1}{2} \max\{1+c_3-(c_1-c_2)(1-p),1+c_3\\+(c_1-c_2)(1-p),
1-c_3-(c_1+c_2)(1-p),\\1-c_3+(c_1+c_2)(1-p)\}-1 .
\end{split}
\end{equation}
Thus we can get the following
\begin{eqnarray}
U&=&2-\frac{1}{2}(1+c_1(1-p))\log_2(1+c_1(1-p))\nonumber\\&&-\frac{1}{2}(1-c_1(1-p))\log_2(1-c_1(1-p)) \nonumber \\
&&-\frac{1}{2}(1+c_2(1-p))\log_2(1+c_2(1-p))\nonumber\\&&-\frac{1}{2}(1-c_2(1-p))\log_2(1-c_2(1-p)), \\
U_{b1}&=&S(\varrho_{AB}), \\
U_{b2}&=&S(\varrho_{AB})+\max\{0,DC\},\\
U_{b3}&=&2S(\varrho_{AB})-2+2(I[\varrho_{AB}(p)]-C[\varrho_{AB}(p)]),\\
\end{eqnarray}
where $DC=\frac{1}{4}(1+c_1+c_2-c_3-c_1p-c_2p)\log_2(1+c_1+c_2-c_3-c_1p-c_2p)
+\frac{1}{4}(1-c_1+c_2+c_3+c_1p-c_2p)\log_2(1-c_1+c_2+c_3+c_1p-c_2p)
+\frac{1}{4}(1+c_1-c_2+c_3-c_1p+c_2p)\log_2(1+c_1-c_2+c_3-c_1p+c_2p)
+\frac{1}{4}(1-c_1-c_2-c_3+c_1p+c_2p)\log_2(1-c_1-c_2-c_3+c_1p-c_2p)-(1-c)\log_2(1-c)-(1+c)\log_2(1+c)$.

\begin{figure}
\subfigure[]
{ \begin{minipage}[b]{0.5\textwidth}
\includegraphics[width=1\textwidth]{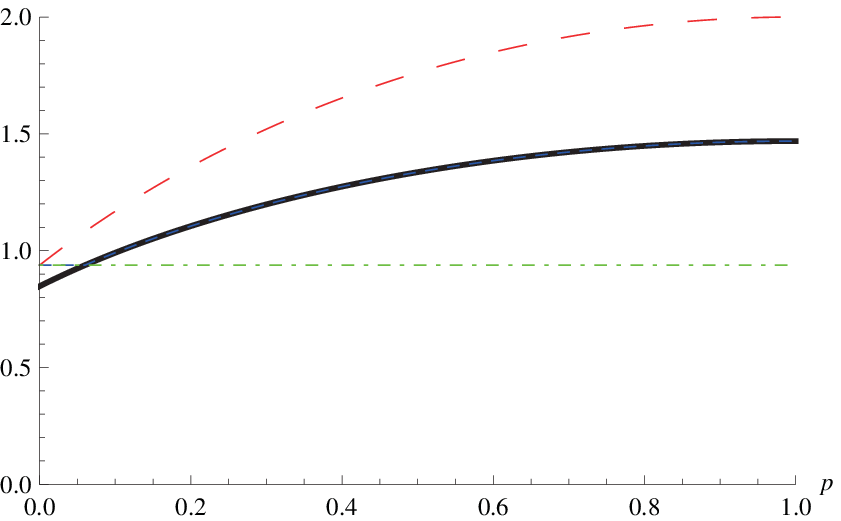}
\end{minipage} }
\subfigure[]
{ \begin{minipage}[b]{0.5\textwidth}
\includegraphics[width=1\textwidth]{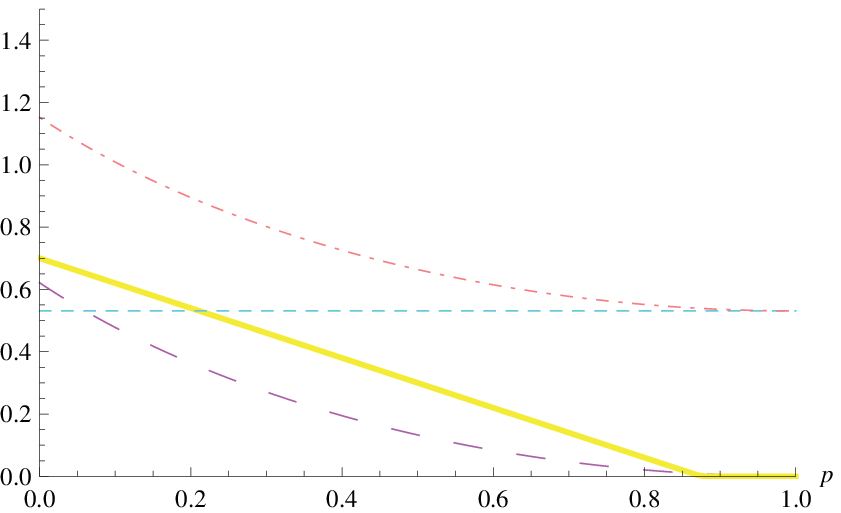}
\end{minipage} }
 \caption{(a) $U$ and $U_{bi} (i=1, 2, 3)$ of the observables $\sigma_1, \sigma_3$ with initial state $c_1=c_2=c_3=-0.8$, under the independently local phasedamping channel with p as the damping rate. $U$-red large dashing color, $U_{b1}$-black solid color, $U_{b2}$-blue dot color, $U_{b3}$-green dot dashed color.
(b) Concurrence (Con)-yellow solid color, discord (D)- magenta large dashing color , classical correlation (C)-light blue dot color,  and mutual information (M)-pink dot dashed color.
 }
\label{fig:Fig2}
\end{figure}

As depicted in Fig.\ 2, choosing $c_1=c_2=c_3=-0.8$, while $U$, $U_{b1}$ and $U_{b2}$ increase all the time while $U_{b3}$ is constant as this unital channel induces the loss of entanglement and discord gradually in asymptotic time, whereas classical correlations remain unchanged. $U_{b1}$ and $U_{b2}$ are almost the same curve during this process, except for a short initial time interval where $U_{b3}$ and $U_{b2}$ coincide. 
This situation should be contrasted with the analysis carried out in Ref. [11], where the behaviours of memory-assisted entropic uncertainty relations with under noise acting on the system qubit are shown. Obvious differences on the behaviours of the uncertainties are due to the different dynamics of quantum and classical correlations of the joint qubits with respect to single qubit decoherence. Furthermore, here we pay more attention to the action of discord-assited memories, while Ref. [11] mainly focuses on the effect of entanglement assistance.

\subsection{Uncertainty relations in non-Markovian environments}

In this section we study the effect of dissipation on the uncertainty relation by exactly solving a model consisting of two independent qubits subject to two zero-temperature non-Markovian reservoirs. We shall see how the behaviour of the uncertainty relation due to correlation dynamics is affected by the environment being, respectively, `quantum' or `classical', {\em i.e.} with or without back-action on the system.

\subsubsection{Reservoirs with system-environment back-action}

The non-Markovian effects on the dynamics of entanglement and discord presented in a two qubits system have been studied recently [19,20]. Assuming a two qubits system A and B whose dynamics is described by the damped Jaynes-Cummings model, the qubits are coupled to a single cavity mode, which in turn is coupled to a non-Markovian environment. The environments are described by a bath of harmonic oscillators, and the spectral density is written as \cite{H. P.}
\begin{equation}
J(\omega)=\frac{1}{2\pi}\frac{\gamma_0\tau^2}{(\omega_0-\omega)^2+\tau^2} ,
\end{equation}
where $\tau$ is associated with the reservoir correlation time $t_B$ by the relation $t_B\approx\frac{1}{\tau}$, and $\gamma_0$ is connected to the time scale $t_{R}$ over which the two-qubit system changes, here $t_R\approx\frac{1}{\gamma_0}$, and the strong coupling condition $t_R<2t_B$ is assumed. The two-qubit Hamiltonian under independent amplitude-damping channels can be written as \cite{F. F. Fanchini}
\begin{equation}
H=\omega_{0}^{j}\sigma_{\dag}^{j}\sigma_{-}^{j}+\sum\limits_{k}\omega_{k}^{j}a_{k}^{(j)\dag}a_{k}^{j}+(\sigma_{\dag}^{j}B^{j}+\sigma_{-}^{j}B^{j\dag}),
\end{equation}
where $B^((j))=\sum\limits_k g_k^((j))a_k^((j))$ with $g_k^((j))$ being the coupling constant, $\omega_0^((j))$ is the transition frequency of the $j^{\rm th}$ qubit, and $\sigma_{\pm}^((j))$ are the system raising and lowering operators of the $j$th qubit. Here the index $k$ labels the reservoir field modes with frequencies$\omega_k^((j))$, and $a_k^((j)^\dag) (a_k^((j)))$ is their creation (annihilation) operator. Here, and in the following, the Einstein convention sum is used. The initial state of the two qubits is the Bell-like state
\begin{equation}
|\psi\rangle=\alpha|00\rangle+\sqrt{1-\alpha^2}|11\rangle .
\end{equation}

According to the dynamics of the initial state's density matrix elements given in Ref. \cite{B. Bellomo}, the mutual information, classical correlation and concurrence present in qubits $A$ and $B$ are given by
\begin{eqnarray}
\fl I[\varrho_{AB}(t)] = -2a^2p_t\log_2(a^2p_t)-2(1-a^2p_t)\log_2(1-a^2p_t)+2a^2p_t(1-p_t)\log_2(a^2p_t(1-p_t)) \nonumber\\
+[-a^2p_t(1-p_t)\nonumber\\
+\frac{1}{2}(1-\sqrt{1-4a^2(1-p_t)p_t})] \log_2(-a^2p_t(1-p_t)+ \frac{1}{2}(1-\sqrt{1-4a^2(1-p_t)p_t})) \nonumber\\
+[-a^2p_t(1-p_t)\nonumber\\
+\frac{1}{2}(1+\sqrt{1-4a^2(1-p_t)p_t})] \log_2(-a^2p_t(1-p_t)+\frac{1}{2}(1+\sqrt{1-4a^2(1-p_t)p_t})),\nonumber\\
\fl D[\varrho_{AB}(t)]  =  \min{D_1,D_2},\nonumber\\
\fl Con(p)= 2\max\{a^2(1-p_t)p_t,\sqrt{2a^2p_t^2+a^4p_t^2(p_t^2-2p_t-1)-2\sqrt{a^4(1-a^2)p_t^4(1-p_ta^2(2-p_t))}}\nonumber\\
 \sqrt{2a^2p_t^2+a^4p_t^2(p_t^2-2p_t-1)+2\sqrt{a^4(1-a^2)p_t^4(1-p_ta^2(2-p_t))}}\}\nonumber\\
 -2a^2(1-p_t)p_t-\sqrt{2a^2p_t^2+a^4p_t^2(p_t^2-2p_t-1)-2\sqrt{a^4(1-a^2)p_t^4(1-p_ta^2(2-p_t))}}\nonumber\\
 -\sqrt{2a^2p_t^2+a^4p_t^2(p_t^2-2p_t-1)+2\sqrt{a^4(1-a^2)p_t^4(1-p_ta^2(2-p_t))}},\nonumber
\end{eqnarray}
where 
\begin{eqnarray}
\fl D_1 = a^2p_t(1-p_t)\log_2(a^2p_t(1-p_t)) \nonumber\\
+[-a^2p_t(1-p_t)+\frac{1}{2}(1-\sqrt{1-4a^2(1-p_t)p_t})]\log_2(-a^2p_t(1-p_t) \nonumber\\
+\frac{1}{2}(1-\sqrt{1-4a^2(1-p_t)p_t}))\nonumber\\
+[-a^2p_t(1-p_t)+\frac{1}{2}(1+\sqrt{1-4a^2(1-p_t)p_t})]\log_2(-a^2p_t(1-p_t)\nonumber\\
+\frac{1}{2}(1+\sqrt{1-4a^2(1-p_t)p_t}))
-a^2p_t\log_2(a^2p_t)-(1-a^2p_t)\log_2(1-a^2p_t)\nonumber\\
-a^2p_t\log_2p_t-a^2p_t(1-p_t)\log_2(1-p_t)-
a^2(1-p_t)p_t\log_2\frac{a^2p_t(1-p_t)}{1-a^2p_t}-\nonumber\\
(1-2a^2p_t+a^2p_t^2)\log_2\frac{1-2a^2p_t+a^2p_t^2}{1-a^2p_t} , \nonumber
\end{eqnarray}
\begin{eqnarray}
\fl D_2=a^2p_t(1-p_t)\log_2(a^2p_t(1-p_t))\nonumber\\
+ [-a^2p_t(1-p_t)+\frac{1}{2}(1-\sqrt{1-4a^2(1-p_t)p_t})]\log_2(-a^2p_t(1-p_t)\nonumber\\
+\frac{1}{2}(1-\sqrt{1-4a^2(1-p_t)p_t}))\nonumber\\
+[-a^2p_t(1-p_t)+\frac{1}{2}(1+\sqrt{1-4a^2(1-p_t)p_t})]\log_2(-a^2p_t(1-p_t)\nonumber\\
+\frac{1}{2}(1+\sqrt{1-4a^2(1-p_t)p_t}))\nonumber\\
-\frac{1+\sqrt{1-4a^2p_t(1-p_t)}}{2}\log_2\frac{1+\sqrt{1-4a^2p_t(1-p_t)}}{2}\nonumber\\
-\frac{1-\sqrt{1-4a^2p_t(1-p_t)}}{2}\log_2\frac{1-\sqrt{1-4a^2p_t(1-p_t)}}{2}. \nonumber
\end{eqnarray}

\begin{figure}
\subfigure[]
{ \begin{minipage}[b]{0.5\textwidth}
\includegraphics[width=1\textwidth]{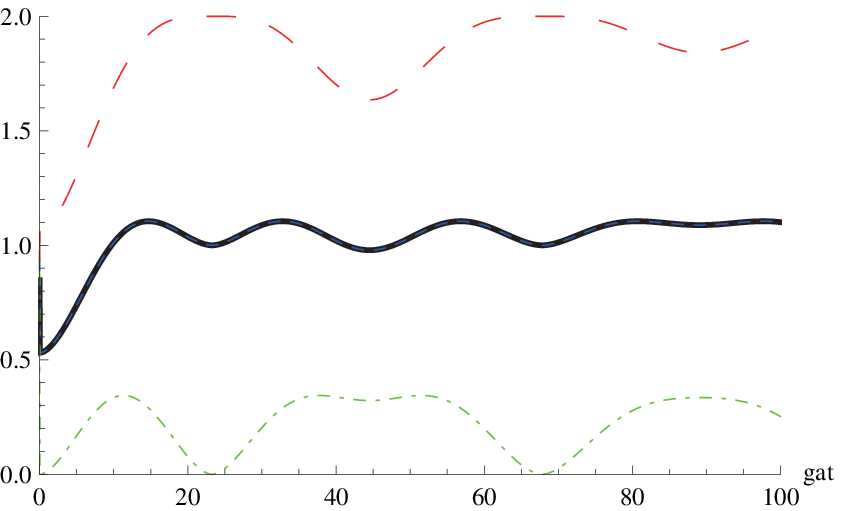}
\end{minipage} }
\subfigure[]
{ \begin{minipage}[b]{0.5\textwidth}
\includegraphics[width=1\textwidth]{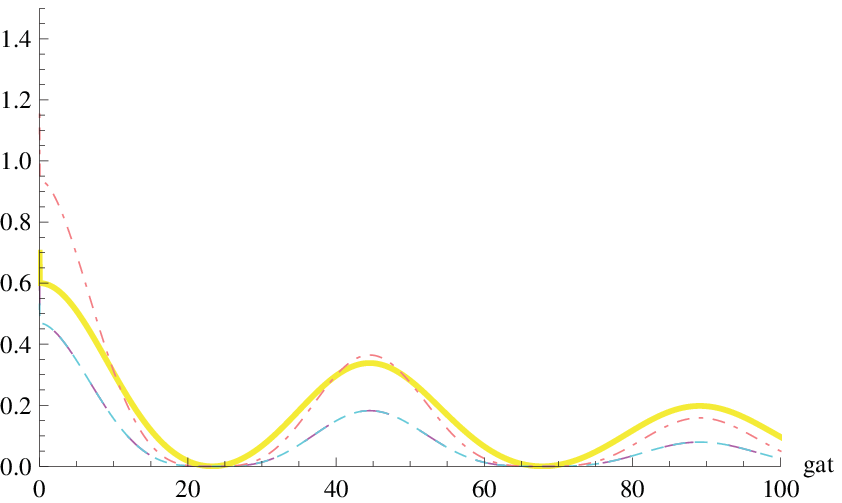}
\end{minipage} }
 \caption{(a) $U$ and $U_{bi} (i=1, 2, 3)$ of the observables $\sigma_1, \sigma_3$ with initial state $|\alpha|00\rangle+\sqrt{(1-\alpha^2 )}|11\rangle $ ($\alpha=\frac{1}{\sqrt{10}}$) under the independently local non-Markovian quantum environments with $\tau=0.01\gamma_0$. $U$-red large dashing color, $U_{b1}$-black solid color, $U_{b2}$-blue dot color, $U_{b3}$-green dot dashed color.
(b) Concurrence (Con)-yellow solid color, discord (D)- magenta large dashing color , classical correlation (C)-light blue dot color,  and mutual information (M)-pink dot dashed color.
 }
\label{fig:Fig3}
\end{figure}

In Fig.\ 3 we plot the uncertainty $U$, as well as the lower bounds $U_{b1}$, $U_{b2}$ and $U_{b3}$ as functions of the rescaled time 
$\gamma_0 t$ in the strong coupling regime, with $1/t=0.01\gamma_0$ and $\alpha=\frac{1}{\sqrt{10}}$. $U$, $U_{b1}$, $U_{b2}$ and $U_{b3}$ all oscillate in the long-time limit due to the entanglement and discord between qubits A and B periodically vanishing and reviving. 
It is apparent that the behaviours of $U_{b1}$ and $U_{b2}$ are the same at short times and tighten the lower bound of the uncertainty with respect of $U_{b3}$. The amplitudes of the oscillations of $U_{b1}$ (or $U_{b2}$) and $U_{b3}$ reduce slowly
as the peaks of entanglement and discord dwindle after each revival.

\subsubsection{Reservoirs without system-environment back-action}

We now want to explore how the entropic uncertainty relations are affected by revivals of correlations, including quantum discord and entanglement, occurring in `classical' non-Markovian environments with no back-action.
Suppose the pair of non-interacting qubits is in a generic initial Bell-diagonal state:
\begin{equation}
\varrho(0)=\sum\limits_{kn}c_{k}^{n}(0)|k^n\rangle\langle k^n|,(k=1,2;n=\pm) ,
\end{equation}
where $|1^{\pm}\rangle=\frac{|01\rangle\pm|10\rangle}{\sqrt{2}}$,$|2^{\pm}\rangle=\frac{|00\rangle\pm|11\rangle}{\sqrt{2}}$. Each of the two qubits is coupled to a random external field acting as a local environment, 
so the global dynamical map $\Omega$ applied on the initial state $\varrho(0)$ is of the random external field type \cite{R. Alicki}
and yields the state
\begin{equation}
\varrho(t)=\frac{1}{4}\sum\limits_{j,k=1}^2U_{j}^{A}(t)U_{k}^{B}(t)\varrho(0)U_{j}^{A\dag}(t)U_{k}^{B\dag}(t)
\end{equation}
where $U_j^R (t)=e^{-iH_j/\hbar}  (R=A,B;j=1,2)$ is the time evolution operator with $H_j=i\hbar g(\sigma_+ e^{-i\phi_j }-\sigma_- e^{i\phi_j })$, and $H_j$ is expressed in the rotating frame at the qubit-field resonant frequency 
$\omega$. In the basis ${|1\rangle, |0\rangle}$, the time evolution operators $U_j^R (t)$ have the following matrix form \cite{R. Lo Franco}
\begin{equation}
U_{j}^{R}(t)=\left(
\begin{array}{cc}
\cos(gt) & e^{-i\phi_j}\sin(gt) \\
e^{-i\phi_j}\sin(gt) &\cos(gt) \\%
\end{array}%
\right),
\end{equation}
where $j=1, 2, \phi_j$ is the phase of the field at the location of each qubit and is either 0 or$\pi$ 
with probability $p = \frac{1}{2}$. The interaction between each qubit and its local field mode is assumed to be strong enough so that, for sufficiently long times, the dissipation effects of the vacuum radiation modes on the qubit dynamics can be neglected. From the matrix of Eq.\ (30) we know that the dynamics is cyclic, and the global map $\Omega$ acts within the class of Bell-diagonal states (or states with maximally mixed marginals \cite{S. Luo}). 
These properties allow us to analytically calculate the correlation quantifiers under the map of Eq.\ (29) for different initial states $\varrho(0)$. 

As shown in Fig.\ 4a (right hand side), for an initial Bell-diagonal state with $c_1^+ (0)=0.9$, $c_1^- (0)=0.1$, and $c_2^+ (0)=c_2^- (0)=0$, choosing $p_1=1-p_2=0.025$, under the map of Eq. (29), both entanglement and classical correlations will collapse and revive during the dynamics, while discord keeps 
approximately constant. 
In Fig.\ 4a (left hand side) we observe that the lower bounds $U_{b1}$ and $U_{b2}$ coincide all the time, and that $U$, $U_{b1}$, $U_{b2}$ and $U_{b3}$ present periodic oscillations: interestingly, $U$ oscillates is out of phase, periodically 
saturating the entanglement-assisted uncertainty relation .
The amplitudes of oscillating revival do not decay. Also, $U_{b1}$ (or the equivalent $U_{b2}$) 
provides one with a tighter lower bound for the uncertainty than $U_{b3}$ at all times. 

\begin{figure}
\subfigure[]
{ \begin{minipage}[b]{0.25\textwidth}
\includegraphics[width=1\textwidth]{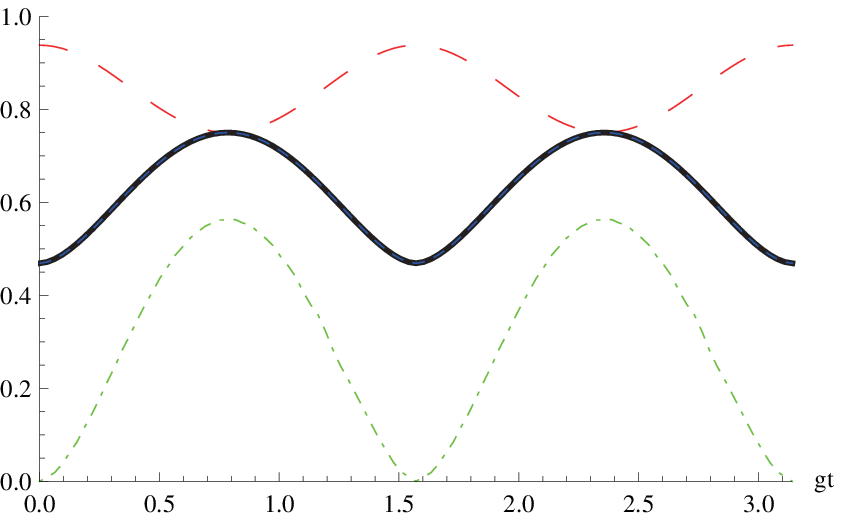}
\end{minipage}
\begin{minipage}[b]{0.25\textwidth}
\includegraphics[width=1\textwidth]{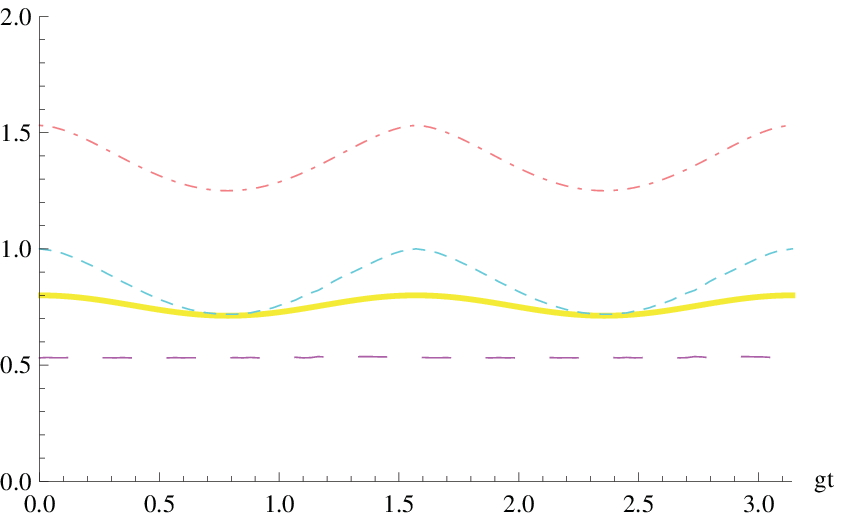}
\end{minipage} }
\subfigure[]
{ \begin{minipage}[b]{0.25\textwidth}
\includegraphics[width=1\textwidth]{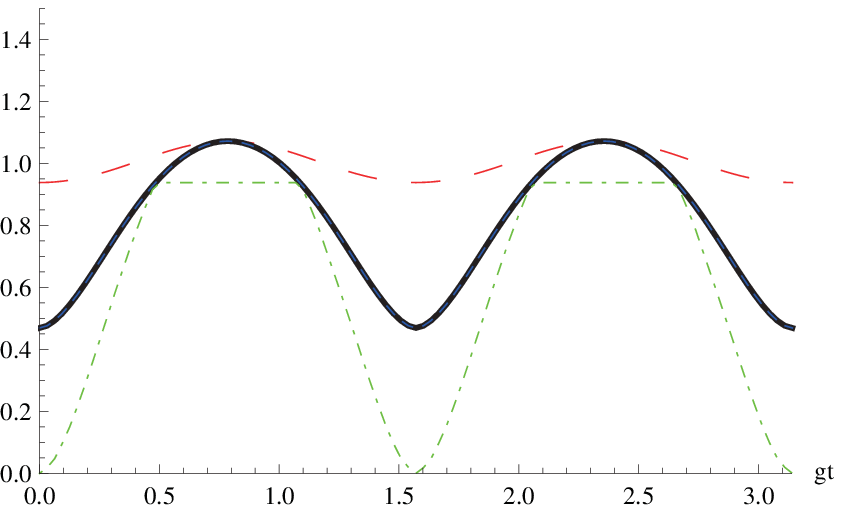}
\end{minipage}
\begin{minipage}[b]{0.25\textwidth}
\includegraphics[width=1\textwidth]{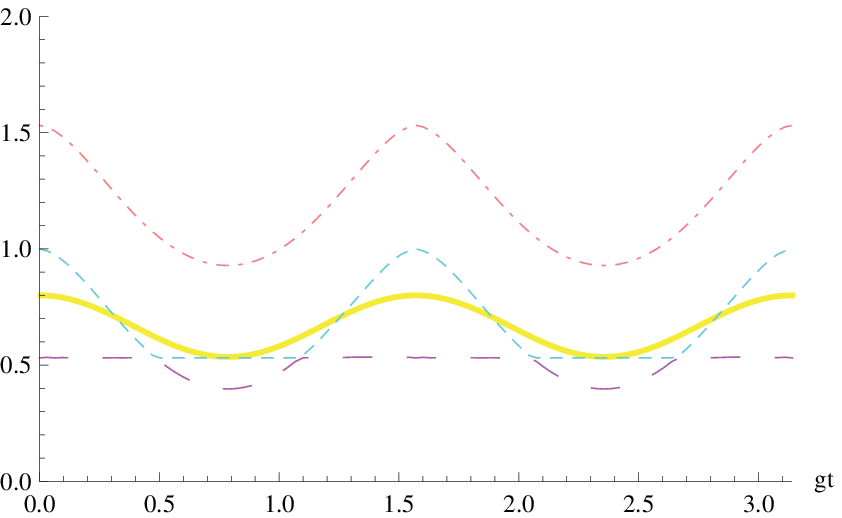}
\end{minipage} }
 \caption{(a) The upper-left figure:$U$ and $U_{bi} (i=1, 2, 3)$ of the observables $\sigma_1, \sigma_3$ with initial state $c_1^+ (0)=0.9$, $c_1^- (0)=0.1$, $c_2^+ (0)=c_2^- (0)=0$ under the independently local non-Markovianc lassical environments with$p_1=0.025$. $U$-red large dashing color, $U_{b1}$-black solid color, $U_{b2}$-blue dot color, $U_{b3}$-green dot dashed color.
The upper-right figure :concurrence (Con)-yellow solid color, discord (D)- magenta large dashing color , classical correlation (C)-light blue dot color,  and mutual information (M)-pink dot dashed color.
(b)The lower-left figure:$U$ and $U_{bi} (i=1, 2, 3)$ of the observables $\sigma_1, \sigma_3$ with initial state as in Fig.4a under the independently local non-Markovian  classical environments  with $p_1=0.08$. $U$-red large dashing color, $U_{b1}$-black solid color, $U_{b2}$-blue dot color, $U_{b3}$-green dot dashed color.
The lower-right figure:concurrence (Con)-yellow solid color, discord (D)- magenta large dashing color , classical correlation (C)-light blue dot color,  and mutual information (M)-pink dot dashed color.
 }
\label{fig:Fig4a}
\end{figure}

The choice $p_1=1-p_2=0.08$ for the same initial state, plotted in Fig. 4b, shows a similar behaviour, with the exceptions that the oscillations of $U$ are now in phase with those of the lower bounds (but the periodic saturation 
of the bounds still occurs), and that $U_{b3}$ provides, at certain points in time, a bound as tight as the other two assisted uncertainty relations.
When entanglement and discord increase to their maximum, $U$, $U_{b1}$ (or $U_{b2}$) and $U_{b3}$ will decrease to their minimum, and viceversa.
It is worth mentioning that the periodical oscillation of the uncertainties and lower bounds 
is only a consequence of the non-Markovian character of the independent qubit-reservoir dynamics, 
whether back-action on the system is present or not. This fact might hence be used to quantify 
the non-Markovianity of single-qubit dynamics.

\subsection{Special cases}

Let us now discuss two special examples of open system dynamics characterised, respectively, by a
particular interplay between quantum discord and classical correlations, and by the presence of discord without 
entanglement.

\subsubsection{Sudden transition between classical and quantum decoherence}

A sharp transition between `classical' and `quantum' loss of correlations in a composite system characterises certain open quantum systems, when properly parametrized. This kind of behaviour has first been noticed in the case of two qubits locally subject to non-dissipative channels \cite{L. Mazzola}, and then observed in an all-optical experimental setup \cite{Jin-Shi}. 
An environment-induced sudden change has also been observed in a room temperature nuclear magnetic resonance setup \cite{R. Auccaise}. Moreover sudden change and immunity against some sources of noise were still found when an environment is modelled as classical instead of quantum \cite{R. Lo Franco}, indicating that such a peculiar behavior is in fact quite general.

Here, we adopt the model of Ref. \cite{L. Mazzola}, supposing the initial state is in the class of states of Eq. (5) and consider two independent phase damping channel, so that the time evolution of the whole system is given by \cite{maziero}
\begin{equation}
\varrho_{AB}(t)=\sum\limits_{kn}\lambda_{k}^{n}(t)|k^n\rangle\langle k^n|,(k=1,2;n=\pm) ,
\end{equation}
where $\lambda_1^{\pm}(t)=\frac{1}{4}(1\pm c_1(t)\mp c_2(t)+c_3(t))$,$\lambda_2^{\pm}(t)=\frac{1}{4}(1\pm c_1(t)\pm c_2(t)-c_3(t))$,
$c_1(t)=c_1 e^{-2\gamma t}$,$c_2(t)=c_2 e^{-2\gamma t}$,$c_3(t)=c_3$, for a damping rate $\gamma$. 
The parameters chosen for the initial state are $c_1=1, c_3=-c_2=0.6$.

 \begin{figure}
\subfigure[]
{ \begin{minipage}[b]{0.5\textwidth}
\includegraphics[width=1\textwidth]{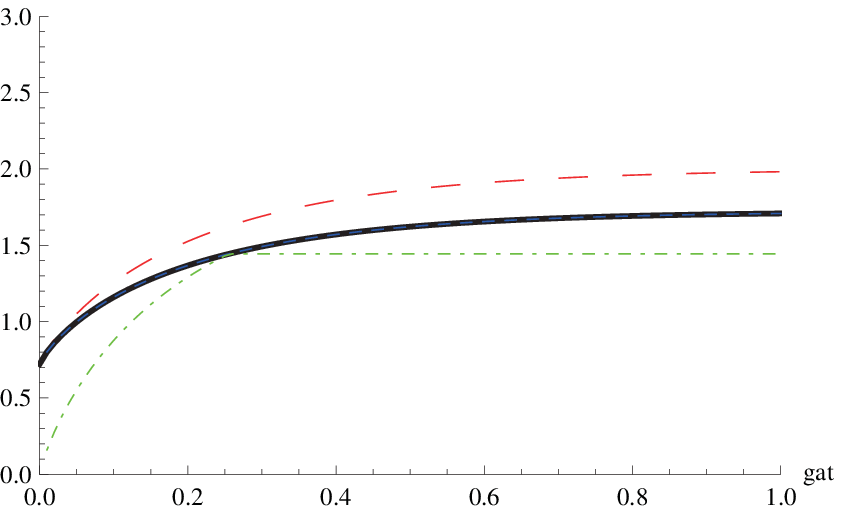}
\end{minipage} }
\subfigure[]
{ \begin{minipage}[b]{0.5\textwidth}
\includegraphics[width=1\textwidth]{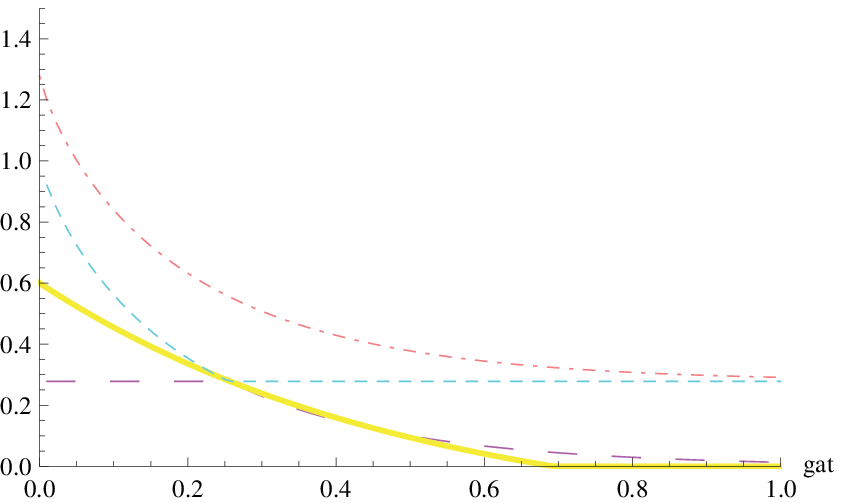}
\end{minipage} }
 \caption{(a) $U$ and $U_{bi} (i=1, 2, 3)$ of the observables $\sigma_1, \sigma_3$ with initial state $c_1=1, c_3=-c_2=0.6$ under the independently local phase damping channel with¦Ãphase damping rate. $U$-red large dashing color, $U_{b1}$-black solid color, $U_{b2}$-blue dot color, $U_{b3}$-green dot dashed color.
(b) Concurrence (Con)-yellow solid color, discord (D)- magenta large dashing color , classical correlation (C)-light blue dot color,  and mutual information (M)-pink dot dashed color.
 }
\label{fig:Fig5}
\end{figure}

As shown in Fig. 5, the uncertainty will increase in the long-time limit due to the gradually missing quantum correlations. The dynamics of correlations are shown on the right plot of Fig. 5. 
In this case, while mutual information and entanglement decrease gradually, 
classical correlations and discord display two mutually exclusive plateaux (when discord remains constant, classical correlation is decreasing, and vice versa). We can observe that this same phenomenon of sudden transition between discord and classical correlation decoherence occurs in the inset of Fig. 4b. Clearly, this sudden transition influences the behaviour of the corresponding entropic uncertainty relations: as shown in Fig. 5, $U_{b1}$ and $U_{b2}$ here coincide all the time (discord does not tighten the entanglement-assisted uncertainty relation) and $U_{b1}$ (or $U_{b2}$) is always a tighter lower bound for the uncertainty than $U_{b3}$. In particular, $U_{b3}$ is increasing at first, and then has a sudden change as its maximum value coincides with $U_{b1}$ (and $U_{b2}$), {\em i.e.}, $U_{b3}$ keeps constant when discord decays in time.

\subsubsection{Quantum correlations without entanglement}

It is well known that many operations in quantum information processing depend largely on quantum correlations 
represented by quantum entanglement. However, there are indications that some protocols might display a quantum advantage without the presence of entanglement \cite{animesh}. Besides, correlations quantified by quantum discord can always be ``activated'', even  
when no quantum entanglement is initially present \cite{marco}. 
We consider here our two qubits system under a one-sided phase damping channel \cite{Jin-Shi}, in the following initial state
\begin{equation}
\begin{split}
\varrho(0)=dR |2^{\dag}\rangle\langle 2^{\dag}|+b(1-R)dR |2^{-}\rangle\langle 2^{-}|+bR |1^{\dag}\rangle\langle 1^{\dag}|+d(1-R)dR |1^{-}\rangle\langle 1^{-}|,
\end{split}
\end{equation}
\begin{figure}
\subfigure[]
{ \begin{minipage}[b]{0.5\textwidth}
\includegraphics[width=1\textwidth]{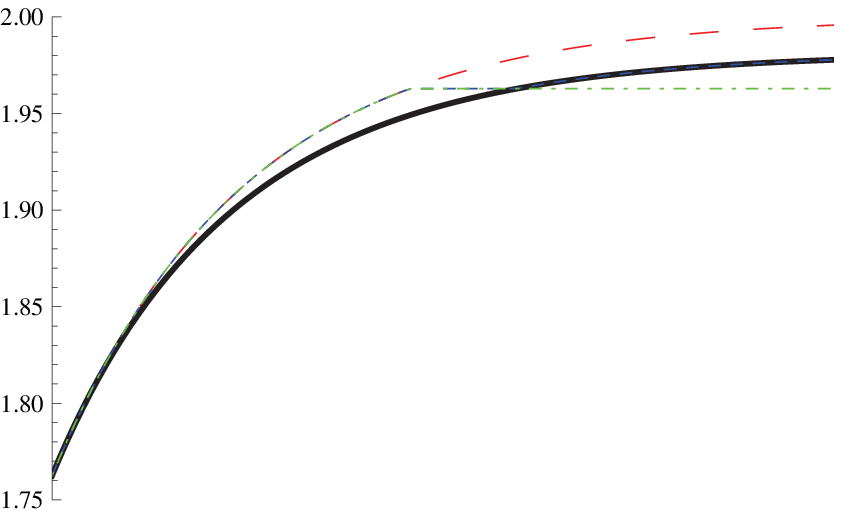}
\end{minipage} }
\subfigure[]
{ \begin{minipage}[b]{0.5\textwidth}
\includegraphics[width=1\textwidth]{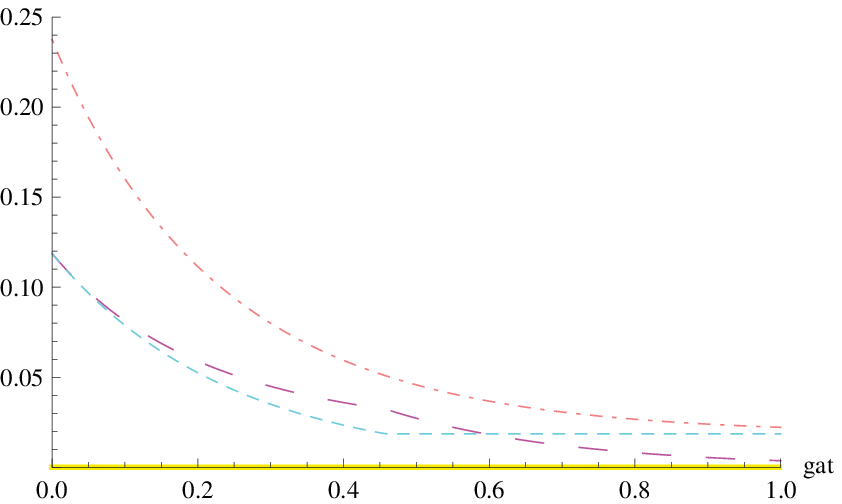}
\end{minipage} }
 \caption{(a) $U$ and $U_{bi} (i=1, 2, 3)$ of the observables $\sigma_1, \sigma_3$ with initial state $b=0.7, R=0.7, d=03$ under a one-sided phase damping channel with damping rate.$U$-red large dashing color, $U_{b1}$-black solid color, $U_{b2}$-blue dot color, $U_{b3}$-green dot dashed color.
(b) concurrence (Con)-yellow solid color, discord (D)- magenta large dashing color , classical correlation (C)-light blue dot color,  and mutual information (M)-pink dot dashed color.
 }
\label{fig:Fig6}
\end{figure}
where $b=0.7, R=0.7$ and $d=0.3$. As shown in the right plot of Fig.6, entanglement is almost zero during this process, while quantum discord is larger than classical correlations for a certain short time interval. We observe the behaviour of the uncertainty relations from Fig. 6: at first, $U$, $U_{b1}$, $U_{b2}$ and $U_{b3}$ coincide and quickly increase; then, while $U$, $U_{b2}$ and $U_{b3}$ coincide (as discord is bigger than classical correlations), $U_{b1}$ is lower than $U$ (or $U_{b2}$). Furthermore, $U_{b1}$ and $U_{b2}$ will provide a higher lower bound than $U_{b3}$ and coincide with each other when discord becomes smaller than classical correlation. 

\section{Conclusions}

We have introduced a new memory-assisted, observable-independent 
entropic uncertainty relation where quantum discord between system and memory plays an explicit role. 
We have shown that this uncertainty relation can be tighter than the ones obtained previously by Berta {\em et al.}\ and Pati {\em et al.}
Moreover, we have explored the behaviour of these three entropic uncertainty relations with assisting quantum correlations for a two-qubit composite system interacting with two independent environments, in both Markovian and non-Markovian regimes. The most common noise channels (amplitude damping, phase damping) were discussed. The entropic uncertainties (or their lower bounds) will increase under independent local unital Markovian noisy channels, while they may be reduced under the non-unital noise channel. The entropic uncertainties (and their lower bounds) exhibit periodically oscillation due to correlation dynamics under independently non-Markovian reservoirs, whether environment is modeled as quantum or classical. In addition, we have compared the differences among three entropic uncertainty relations 
in Eq.\ (\ref{berta}), (\ref{pati}) and (\ref{bound}). The lower bound $U_{b2}$ or $U_{b3}$ will tighten the bound on the uncertainty 
when discord is bigger than classical correlation, which is often the case in practice. 
The relation between quantum correlations and the uncertainties is subtle, since a certain reduction int he uncertainty may also happen in the presence of small quantum correlations without entanglement. 
We have also shown that, in essence due to the greater resilience of the nearly ubiquitous quantum discord \cite{ferraro}, situations arise where uncertainty relations tightened by quantum discord offer a better estimate of the actual uncertainties in play. However, the advantage offered in this sense by the quantity $U_{b3}$ of Eq.\ (\ref{bound}) 
seems to be limited to rather specific circumstances.

\ack
This work was supported by the National Natural Science Foundation of China under Grant 61144006, by the Foundation of China Scholarship Council, by the Project Fund of Hunan Provincial Science and Technology Department under Grant 2010FJ3147, and by the Educational Committee of the Hunan Province of China through the Overseas Famous Teachers Programme.


\medskip



\begin{thebibliography}{99}

\bibitem{Heisenberg} Heisenberg, W., Z. Phys. {\bf 43}, 172 (1927).

\bibitem{Robertson} Robertson, H. P., Phys. Rev. {\bf 34}, 163 (1929).

\bibitem{Deutsch} Deutsch, D., Phys. Rev. Lett. {\bf 50}, 631 (1983).

\bibitem{Kraus}Kraus, K., Phys. Rev. D {\bf 35}, 3070 (1987); Maassen, H. $\&$ Uffink, J. B. M., Phys. Rev. Lett. {\bf 60}, 1103 (1988).

\bibitem{Berta} Berta, M., Christandl, M., Colbeck, R., Renes, J. M. $\&$ Renner, R., Nature Phys. {\bf 6}, 659 (2010).

\bibitem{Li} Li. Chuan-Feng, Jin-Shi Xu, Xiao-Ye Xu, Ke Li, and Guang-Can Guo, Nature Phys. {\bf 7}, 752 (2011); Robert Prevedel, Deny R. Hamel, Roger Colbeck, Kent Fisher, and Kevin J. Resch, Nature Phys. {\bf 7}, 757 (2011).

\bibitem{ved} L. Henderson and  V. Vedral,
J. Phys. A. {\bf 34}, 6899 (2001).

\bibitem {Ollivier}
H. Ollivier and W. H. Zurek, Phys. Rev. Lett. \textbf{88}, 017901 (2001).

\bibitem{Pati} Arun Kumar Pati, Mark M. Wilde, A. R. Usha Devi, A. K. Rajagopal, and Sudha, , arxiv:1204.3803v2.

\bibitem{tomamichel}M. Tomamichel and R. Renner, Phys. Rev. Lett. {\bf 106}, 110506 (2011).

\bibitem{ng} H. Y. Ng, M. Berta, and S. Wehner, Phys. Rev. A {\bf 86}, 042315 (2012).

\bibitem{moloktov} S. N. Moloktov, JETP Letters {\bf 94}, 820 (2012).

\bibitem{hanggi} E. H\"anggi and S. Wehner, arXiv:1205.0842.

\bibitem{guhne} O. G\"uhne and M. Lewenstein, Phys. Rev. A {\bf 70}, 022316 (2004).

\bibitem{niekamp} S. Niekamp, M. Kleinmann, and O. G\"uhne, J. Math. Phys. {\bf 53}, 012202 (2012).

\bibitem{wehner}S. Wehner and A. Winter, New J. Phys. {\bf 12}, 025009 (2010).

\bibitem{Coles}Patrick J. Coles, L. Yu, V. Gheorghiu,  R. B. Griffiths, Phys. Rev. A {\bf 83}, 062338 (2011).

\bibitem{T. Yu} T. Yu and J. H. Eberly, Science {\bf 323}, 598 (2009).

\bibitem{maziero} J. Maziero, L. C. Celeri, R. M. Serra, and V. Vedral, Phys. Rev. A {\bf 80}, 044102 (2009).

\bibitem{Francesco} F. Ciccarello and V. Giovannetti, Phys. Rev. A {\bf 85}, 010102(R) (2012); T. Werlang, S. Souza, F. F. Fanchini, and C. J. Villas Boas, Phys. Rev. A {\bf 80}, 024103 (2009).

\bibitem{Wang} Wang , B . , Xu , Z . - Y . , Chen , Z . - Q . $\&$ Feng , M, Phys. Rev. A {\bf 81} , 014101 (2010); F. F. Fanchini, T. Werlang, C. A. Brasil, L. G. E. Arruda, and A. O. Caldeira, Phys. Rev. A {\bf 81}, 052107 (2010).

\bibitem{Z. Y.} Z. Y. Xu, W. L. Yang, and M. Feng, arxiv:1203.3331.

\bibitem{H. P.} H. P. Breuer and F. Petruccione, The Theory of Open Quantum Systems (Oxford University Press, Oxford, 2002); H. Carmichael, An Open Systems Approach to Quantum Optics (Springer, Berlin, 1993).

\bibitem{M. A. Nielsen} M. A. Nielsen and I. L. Chuang, Quantum Computation and Quantum Information (Cambridge University Press, Cambridge, 2000).

\bibitem{G. Jaeger} G. Jaeger, Quantum Information-An Overview (Springer, New York, 2007).

\bibitem{J. Maziero} J. Maziero, T. Werlang, F. F. Fanchini, L. C. Celeri, and R. M. Serra, Phys. Rev. A {\bf 81}, 022116 (2010).

\bibitem{S. Luo} S. Luo, Phys. Rev. A {\bf 77}, 042303 (2008).

\bibitem{B. Bellomo} B. Bellomo, R. Lo Franco, and G. Compagno, Phys. Rev. Lett. {\bf 99}, 160502 (2007).

\bibitem{F. F. Fanchini} F. F. Fanchini, T. Werlang, C. A. Brasil, L. G. E. Arrud, and A. O. Caldeira, Phys. Rev. A {\bf 81}, 052107 (2010).

\bibitem{R. Alicki} R. Alicki and K. Lendi, Quantum Dynamical Semigroups and Applications, Lect. Notes Phys. 717 (Springer, Berlin Heidelberg, 2007).

\bibitem{R. Lo Franco} R. Lo Franco, B. Bellomo, E. Andersson, and G. Compagno, Phys. Rev. A {\bf 85}, 032318 (2012).

\bibitem{B. Julsgaard} B. Julsgaard, J. Sherson, J. I. Cirac, J. Fiurasek, E. S. Polzik, Nature (London) {\bf 432}, 482 (2004).

\bibitem{L. Mazzola} L. Mazzola, J. Piilo, and S. Maniscalco, Phys. Rev. Lett. {\bf 104}, 200401(2010).

\bibitem{Jin-Shi} Jin-Shi Xu, Xiao-Ye Xu, Chuan-Feng Li, Cheng-Jie Zhang, Xu-Bo Zou and Guang-Can Guo, Nature Commun. {\bf 1}, 7 (2010).

\bibitem{R. Auccaise} R. Auccaise, L. C. Celeri, D. O. Soares-Pinto, E. R. de Azevedo, J. Maziero, A. M. Souza, T. J. Bonagamba, R. S. Sarthour, I. S. Oliveira, and R. M. Serra, Phys. Rev. Lett. {\bf 107}, 140403 (2011).

\bibitem{animesh}A. Datta, A. Shaji, and C. M. Caves, Phys. Rev. Lett. {\bf 100}, 050502 (2008).

\bibitem{marco} M. Piani, S. Gharibian, G. Adesso, J. Calsamiglia, P. Horodecki, and A. Winter, 
Phys. Rev. Lett. {\bf 106}, 220403 (2011); A. Streltsov, H. Kampermann, and D. Bruss, Phys. Rev. Lett.
{\bf 106}, 160401 (2011).

\bibitem{ferraro}A. Ferraro, L. Aolita, D. Cavalcanti, F. M. Cucchietti, and A. Ac\'{\i}n, 
Phys. Rev. A {\bf 81}, 052318 (2010).


\end{thebibliography}
\end{document}